\documentclass[fleqn,10pt]{wlscirep}
\usepackage[utf8]{inputenc}
\usepackage[T1]{fontenc}
\usepackage{amsmath}
\usepackage{wasysym}

\begin{document}

\title{Emergence of scaling in dockless bike-sharing systems} 

\author[1,*]{Ruiqi Li}
\author[1]{Ankang Luo}
\author[1]{Fan Shang}
\author[2,3,9,*]{Linyuan L\"u}
\author[4,*]{Jingfang Fan}
\author[1]{Gang Lu}
\author[5]{Liming Pan}
\author[6,7]{Lixin Tian}
\author[8,*]{H. Eugene Stanley}

\affil[1]{UrbanNet Lab, College of Information Science and Technology, Beijing University of Chemical Technology, Beijing 100029, China}
\affil[2]{Yangtze River Delta Institute, University of Electronic Science and Technology of China, Huzhou 313000, China}
\affil[3]{Institute of Fundamental and Frontier Sciences, University of Electronic Science and Technology of China, Chengdu 610054, China}
\affil[4]{School of Systems Science, Beijing normal University, Beijing 100875, China}
\affil[5]{School of Computer and Electronic Information, Nanjing normal University, Nanjing 210023, China}
\affil[6]{School of Mathematical Sciences, Jiangsu University, Zhenjiang 212013, China}
\affil[7]{School of Mathematical Science, Nanjing normal University, Nanjing 210042, China}
\affil[8]{Center for Polymer Studies and Physics Department, Boston University, Boston, MA 02215, USA}
\affil[9]{Beijing Computational Science Research Center, Beijing 100193, China}
\affil[*]{corresponding authors: lir@buct.edu.cn (Ruiqi Li), linyuan.lv@uestc.edu.cn (Linyuan L\"u), jingfang@bnu.edu.cn (Jingfang Fan), hes@bu.edu (H. Eugene Stanley).} %


\begin{abstract}
\textbf{Fundamental laws of human mobility have been extensively studied, yet we are still lacking a comprehensive understanding of the mobility patterns of sharing conveyances. Since travellers would highly probably no longer possess their own conveyances in the near future, the interplay between travellers and sharing bikes is a central question for developing more sustainable transportation. Dockless bike-sharing systems that record detailed information of every trip provide us a unique opportunity for revealing the hidden patterns behind riding activities. By treating each bike as an individual entity, we reveal that distributions of mobility indicators of bikes are quite different from humans; and mobility patterns are even inconsistent across cities. All above discrepancies can be well explained by a choice model that is characterized by a universal scaling. Our model unveils that instead of choosing among the newest bikes, the distribution of rank values of selected bikes on usage condition manifests a truncated power-law and is quite stable across several cities despite various diversities. Our framework would have broad implications in sharing economy and contribute towards developing a greener, healthier, and more sustainable future city.} 
\end{abstract}

\flushbottom
\maketitle

\section*{Introduction}  
Fundamental laws of human mobility have been extensively studied over past decades, due to great importance to various urban studies, including urban planning and design \cite{batty2013new,xu2018planning,olmos2020data}, predicting the evolution of urban systems \cite{li2017simple}, containing the spreading of infectious disease \cite{viboud2006synchrony,brockmann2013hiddenGeo,li2017effects,li2018effect,jia2020population}, traffic engineering \cite{helbing2001traffic,dong2016population,olmos2018macroscopic}, emergency management \cite{bagrow2011collective,lux2012predictability}, and resource allocations \cite{ruan2020dynamic}. 
Seminal works on revealing the general laws of massive human mobility at a relatively high spatio-temporal granularity and a large scale started only one and a half decades ago. Brockmann et al. \cite{brockmann2006scaling} made quantitative assessments of massive human travel behaviours across multiple spatial scales possible by exploiting the circulation of around 0.5 million banknotes, whose mobility is driven by the travelling and interactions of humans. They discovered that the distributions of travel distance 
exhibit a power-law, which shows that long-distance trips are not rare and take up a considerable fraction. 
With the development of information and communication technologies, massive cellphone data, which collects passive geo-located data of individuals when using services at a fine spatio-temporal resolution, enable us to have more consistent and comprehensive descriptions of human mobility at both population level and individual level. 
Both mobility displacement and radius of gyration at the population level of human movements have been proved to be following a truncated power-law distribution 
by analyzing the trajectories of 100,000 anonymized mobile phone users over six months \cite{gonzalez2008understanding}. 
Based on exploration and preferential return mechanisms, most scaling behaviours of human mobility at both individual and population levels can be derived \cite{song2010modelling}. A simple scaling relation can connect both human mobility and social interactions \cite{deville2016scaling}. Recent discovered scaling law on visitation behaviours predict that the number of visitors to any location decreases as the inverse square of the product of visiting frequency and travel distance \cite{schlapfer2021universal}. 
These simple scaling relations, which are also ubiquitous in other systems \cite{barzel2013universality,sornette2006critical,liA2017fundamental}, significantly increase our ability to understand complex human behaviours and intervening related complex systems \cite{liA2017fundamental}. 

Apart from walking, our mobility largely relies on certain sorts of transportation means, including bikes, cars, trains, and planes. 
Yet, in comparison, we lack a coherent understanding of the mobility patterns of transport conveyances and their relations with travellers. Such problems are nontrivial in the era of sharing economy \cite{sundararajan2016sharing,shaheen2016mobility},  
as a traveller would highly probably no longer possess his/her own conveyance in the near future.   
The answers to both questions would be valuable to optimize the daily operation of ride-sharing platforms to improve the travel experience of users and to save resources allocated in transportation systems \cite{vazifeh2018addressing} and mitigate traffic congestion. On the way towards the transition to a greener, more sustainable and resilient mobility, bike-sharing systems have been regarded as a promising solution \cite{jiang2020dockless} 
to reduce carbon emissions and improve public health \cite{dill2003bicycle,jappinen2013modelling,hull2014bicycle,xu2019unravelBIKE,jiang2020dockless,2019bikelife,cheng2021role} due to its usage flexibility, saving of parking space and fossil energy by replacing short-distance motorized trips, increasing fitness and reducing the stress of riders from cycling activities. Since biking allows for greater social distancing than other means of public transportation \cite{jiang2020dockless}, bike-sharing systems are proved to be more resilient and safer to move around for essential needs during the COVID-19 pandemic \cite{teixeira2020link}. After the pandemic, biking is also expected to be more frequent in cities \cite{2020meninoSurvey}. 

Recently, due to great advances in IoT (Internet of Things) and mobile payment technology, there is a global booming of dockless bike-sharing platforms (including Mobike, Ofo, DiDi Bike, LimeBike, Spin, Ford GoBike). Taking China as an instance, dockless sharing bikes have been quite popular \cite{sun2018sharing}. It has been reported that more than 360 cities have dockless bike-sharing systems, with an average of 47 million trips each day \cite{jiang2020dockless}. 
Comparing with docked station-based sharing bikes, which have to be returned to a dock at certain fixed established sites (and the number of available docks there is usually limited), dockless ones are free from such restrictions and give better accessibility and more flexibility to users \cite{li2021gravity}. 
A user does not need to carry his/her own bike anymore and can pick up an available dockless sharing bike nearby, which is similar to Zipcar and far less constrained by the available picking-up/dropping-off locations.  
Users can park and ``return'' a dockless sharing bike anywhere suitable for a bike near the destination. 
Thus, the data presented in dockless bike-sharing systems is much closer to all potential biking travel demands and natural riding dynamics. 
Better managed dockless bike-sharing systems can hopefully bring a myriad of health, climate, and economic benefits to cities, reshaping urban mobility towards a more sustainable track and giving urban dwellers a more resilient and effective way to travel \cite{jiang2020dockless}. 

Yet, there are many challenges of maintaining the efficiency of the system, controlling the cost, and keeping sidewalks and other urban public spaces neat and friendly. As bikes are motorless, resolving the above problems strongly depends on a better understanding of mobility patterns of sharing conveyances driven by the heterogeneous spatio-temporal travel demands of riders, and more importantly, depends on the relation between travellers and sharing conveyances -- the choice made by users on choosing which bike to ride. 
By recording detailed information of every trip due to billing purpose, dockless bike-sharing systems provide us a unique opportunity for approaching solutions to such problems.
The dockless bike-sharing platform usually records the departure and arrival location as well as the start and end timestamp of each trip (and in some cases, even pass-by points of the whole riding trajectory), and more importantly, the anonymized identity of users and bikes, which was impossible for previous studies on biking behaviours.

In this work, by exploiting unique datasets of seven diversified cities across countries obtained from big dockless bike-sharing platforms, including \textit{Mobike}, \textit{DiDi Bike}, and \textit{Hellobike} (see Methods and Supplementary Tables 1-2 for more details of datasets),
we investigate the mobility patterns of both riders and bikes, riding dynamics, and the relation between riders and sharing conveyances. By treating each bike as an individual entity, we reveal that the mobility patterns of bikes at the population level and individual level can be quite different from humans on various indicators. 
And more unsettling, some mobility patterns can be inconsistent across cities, which indicates that there might be no universal regularities behind riding behaviours.  
But reassuringly, we discover that all above discrepancies can be reasonably well explained by a universal scaling behaviour that emerges from the interplay between travellers and bikes (i.e., choices made by users on choosing which bike to ride). Despite various diversities across cities and over time, including different user-bike ratios, dissimilar characteristics of road networks and urban terrain, climate and weather, or even different attitudes towards biking or level of curiosity on sharing bikes, we find that such a scaling behaviour is quite robust.  
The discovery of such a universal scaling relation  
allows us to make better predictions on the bikes that a user chose to ride, which will be
valuable to design a more efficient and user-friendly dockless bike-sharing system, as well as to its sustainable operation.

\section*{Results}
\subsection*{A glance at biking behaviours in cities.}
To study mobility patterns of riders and sharing conveyances, and the relation between users and bikes in complex dockless bike-sharing systems, we collect seven large-scale datasets recording more than ten million riding trips across seven highly diversified cities across countries -- Shanghai (D1), Beijing (D2), Nanjing (D3-D4, obtained from different platforms), Chengdu (D5), Xi'an (D6), Xiamen (D7) in China, and Singapore (D8) -- with high spatio-temporal resolution (see brief summary in Supplementary Tables 1-2 and Supplementary Note 1 for more details). These cities are far from each other and are located in different regions with diversified demographic and socioeconomic statuses, urban geography, climate, and regional custom. 
The datasets also span several years, from 2016 to 2020. 

\begin{figure}[ht] \centering
  \includegraphics[width=\linewidth]{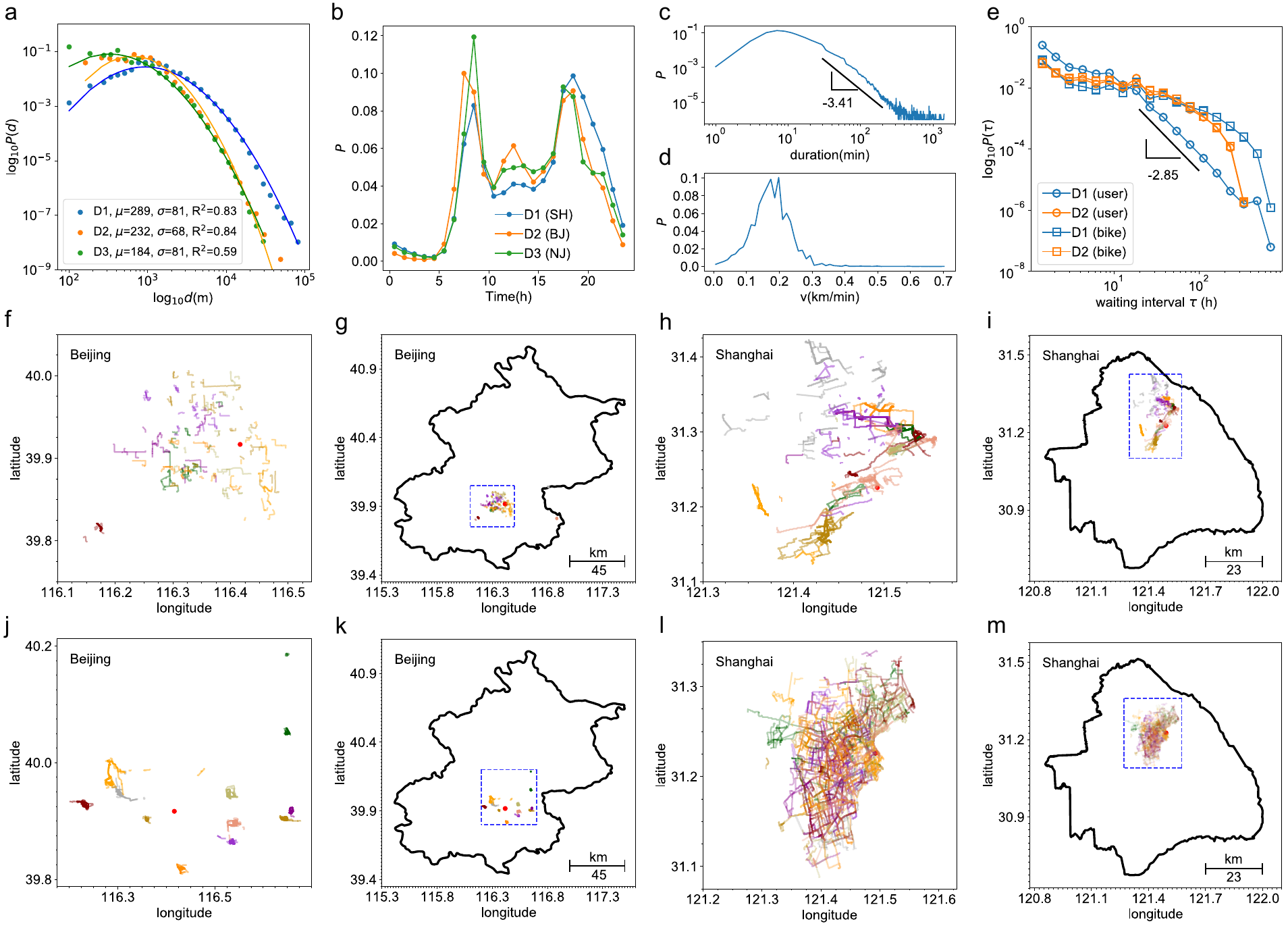}
  \caption{Statistical features of riding dynamics, and trajectories of the ten most active riders and bikes in Beijing and Shanghai. 
  \textbf{a}, the distribution of travel distance of biking trips, which can be fitted by a log-normal function.  
  Note that for D2 and D3, no realistic trajectory are provided, thus the trip distance is estimated from the routing path in the road network that connects the origin and the destination, which is obtained from Amap API (\url{https://lbs.amap.com/}, see Supplementary Fig. 1, Supplementary Table 2, and Supplementary Note 1 for more details). 
  \textbf{b}, the average fraction of riding activities by the time of a day, which is generally of three peaks across cities. In comparison, Beijing and Nanjing are of a higher noon peak and a higher morning peak than evening. While in Shanghai, it is the opposite, and the evening peak is wider, which might indicate that bikes are used for other activities besides commuting. 
  \textbf{c}, the distribution of riding duration and \textbf{d}, average riding speed of each trip in Shanghai. The riding duration is fat-tailed, while the riding speed is more concentrated and closer to a normal distribution. 
  \textbf{e}, the distribution of waiting interval between two consecutive riding activities of riders (marked as {\scriptsize $\Circle$}) and bikes ({\scriptsize $\square$}).  
  The distribution of riders in Shanghai is closer to a power-law, while the case in Beijing 
  is flatter. The distribution of bikes in Beijing happens to be almost identical to riders' and is similar to the case in Shanghai. Lines without a fitting exponent in (\textbf{b, e}) are guidance to eyes.  
  \textbf{f-i}, Trajectories of the ten most active (\textbf{f-g}) bikes and (\textbf{j-k}) riders in Beijing and  (\textbf{h-i, l-m}) Shanghai. Each entity (either a rider or a bike) is denoted by a different colour, the position of trajectories relative to the whole city is shown on the right side. The red dots in (\textbf{f-i}) indicate the city centre, and the left-hand side figures are a zoom-in of the blue dashed regions in the right-hand side figures.}
 \label{fig:basics_trajectories}
\end{figure}

We first analyse some statistical features of riding dynamics. The distributions of displacement distance of all riding trips across cities can be better fitted by log-normal distributions (see Fig.\ref{fig:basics_trajectories}a). 
The difference across cities might originate from different urban terrain and partially from the way of calculating displacement distance (see Methods). 
The average fraction of riding activities by the time of a day is generally of three peaks at rush hours and at noon, which indicates that bikes are mainly used during commuting and lunch-related trips (see Fig.~\ref{fig:basics_trajectories}b).  
The distribution of riding duration in Shanghai is fat-tailed (see Fig.~\ref{fig:basics_trajectories}c, and Supplementary Fig.~1c for other cities, which also exhibit a long tail but might be subject to artificial truncation). This indicates that the average riding duration may not be that representative as ordinarily assumed \cite{chen2019analyzing}. 
Since we also have the biking trajectory in the Shanghai dataset (D1), we can make a relatively accurate calculation on the average riding velocity of each trip, the distribution of which is closer to a normal distribution (see Fig.~\ref{fig:basics_trajectories}d). 

Moreover, we also analyse the distribution of inter-trip waiting intervals between two consecutive rides of users (see Fig.~\ref{fig:basics_trajectories}e), 
which, in Shanghai (D1), roughly manifests a bursty pattern (depicted by a power-law) similar to the ones observed on a variety of human behaviours \cite{barabasi2005origin}, indicating that although most consecutive biking trips are made soon after a previous one, occasionally there are long periods without any biking activity. When treating bikes as individual entities, we can also obtain the corresponding distribution for bikes, which 
differs from the distribution of users in Shanghai. 

While the distributions of both bikes and users in Beijing (D2) are quite similar
(see Fig.~\ref{fig:basics_trajectories}e). In addition, such patterns are not that consistent across cities (see Supplementary Fig.~1d for the distributions of bikes in other cities, which are closer to a power-law).   

The trajectories of the ten most active users and bikes (i.e., the ones with the largest number of trips) in Shanghai (D1) and Beijing (D2) also reveal some inconsistent patterns (see Fig.\ref{fig:basics_trajectories}f-i and more details in Supplementary Note 1 for estimating the trajectory in the Beijing dataset). 
In Shanghai, we observe that trajectories of the ten most active users span a much wider range (see Fig.\ref{fig:basics_trajectories}l-m), while the ones of bikes are relatively concentrated (see Fig.\ref{fig:basics_trajectories}h-i). However, in Beijing, we observe an almost opposite pattern: the most active users are riding within a small region (see Fig.\ref{fig:basics_trajectories}j-k), while bikes are covering a broader area (see Fig.\ref{fig:basics_trajectories}f-g). The most active users in Beijing might use dockless sharing bikes as a commuting connector, while those a few most active users in Shanghai might ride them to explore the city. As for bikes, the different patterns might be originated from their interplay with users, as well as the daily rebalancing operations conducted by the dockless bike-sharing company.
It is worth noting that the trajectories of bikes are not necessarily connected due to daily operations. We find that roughly 2/3 of bikes had been moved from their last arrival location in both Shanghai (D1) and Beijing (D2), while in Nanjing (D3), this fraction reaches 83\%; in Singapore (D8), it is around 34\%  (see Supplementary Fig. 2).

\subsection*{Inconsistent mobility patterns of travellers and sharing conveyances across cities}
The trajectories of those most active travellers and bikes give us the impression that the mobility patterns of users and bikes might be different and even inconsistent across cities.  
To make quantitative comparisons, we investigate various key features of mobility patterns at both individual and population levels.  
\subsubsection*{Individual level}

\begin{figure}[ht] \centering
  \includegraphics[width=\linewidth]{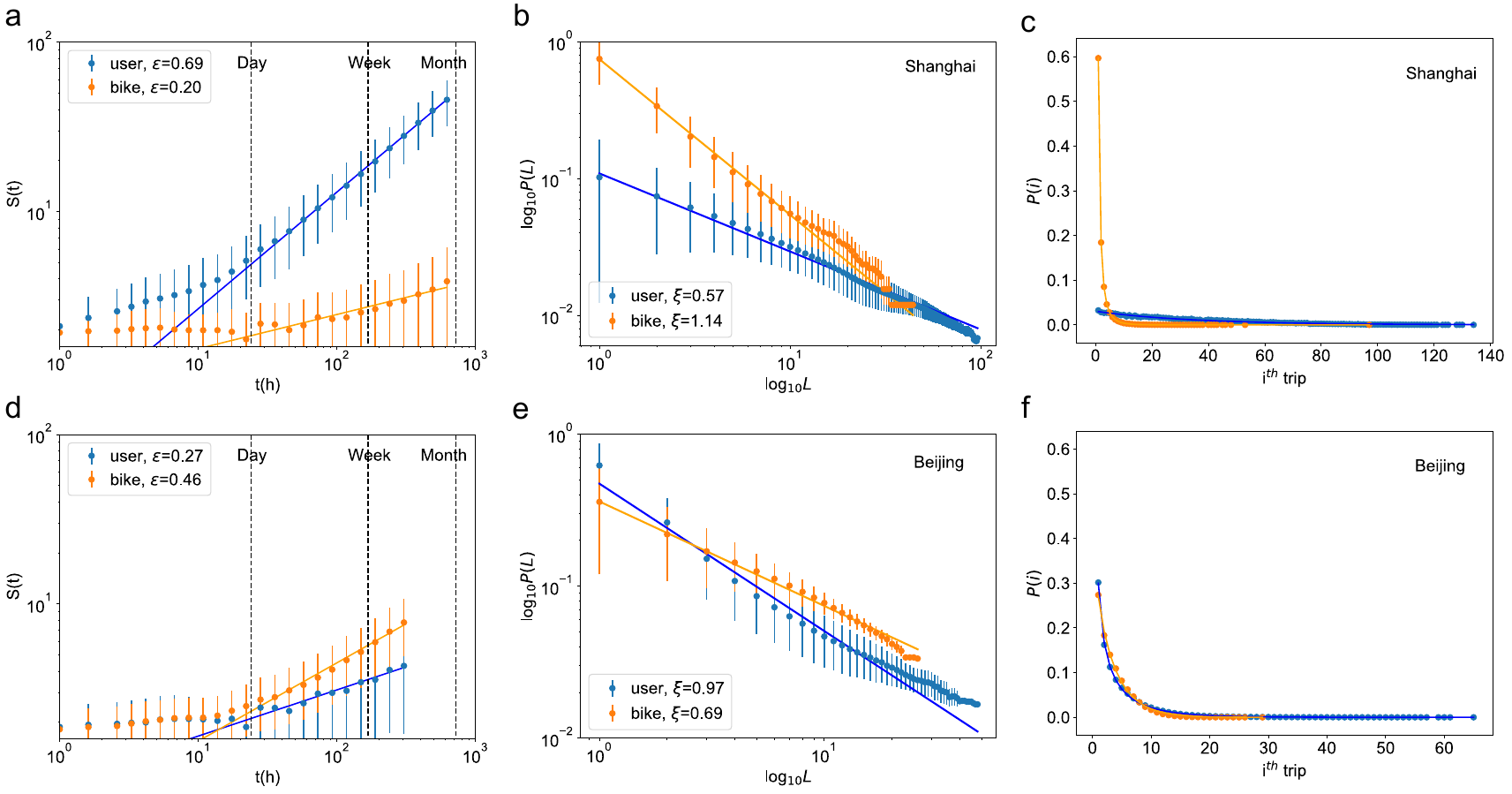} 
  \caption{Mobility patterns of riders and bikes at the individual level in Shanghai (\textbf{a-c}) and Beijing (\textbf{d-f}). 
  \textbf{a, d}, the exploration of new locations of riders and bikes along with the increase of observation duration $t$, which are qualitatively consistent but exhibiting opposite patterns across Shanghai (\textbf{a}) and Beijing (\textbf{d}). The users in Shanghai explore more unique locations than bikes over a month with a larger scaling exponent (0.69 and 0.20 for users and bikes, respectively), which is exactly the opposite in Beijing over two weeks (0.27 and 0.46 for users and bikes, respectively). 
  \textbf{b, e}, the revisitation dynamics on locations of users and bikes. They are again qualitatively the same, but the scaling exponents are different and manifest opposite patterns across cities.  
  In Beijing (\textbf{e}), riders have a stronger tendency to revisit former favored locations indicated by a deeper slope -0.97 (against -0.69 for bikes), which is also qualitatively consistent with results in Fig. 1j-k; while in Shanghai (\textbf{b}), the slopes are -0.57 and -1.14 for users and bikes, respectively, which again exhibits an opposite pattern across cities. We indeed observe a stronger tendency of exploration for those most active users than bikes in Shanghai, and vice versa in Beijing (see Fig. 1f-m). 
  \textbf{c, f}, the distribution of the longest trip of users and bikes. The longest trip of a bike tends to be highly concentrated in first a few trips, 
  while the case for users is quite uniform in Shanghai (\textbf{c}); In Beijing (\textbf{f}), both distributions are quite similar. Error bars mean $\pm$sd.  
  }
 \label{fig:mobility-individual}
\end{figure}

We first rasterize the whole urban space into 1 km $\times$ 1 km grids, each of which is identified as a unique location. Then we compare the exploration dynamics of both riders and bikes in Shanghai (see Fig. \ref{fig:mobility-individual}a). Along with the increase of observation duration, the number of explored unique locations $S(t)$ grows sublinearly for both riders and bikes $S(t)\propto t^\epsilon$, where $t$ is the observation duration of each entity since its first trip. This pattern is also temporally consistent (see Supplementary Fig. 3) and is qualitatively in agreement with previous findings \cite{song2010modelling}. The scaling exponent of users ($\epsilon_{Beijing}^{user}$=0.69) is substantially larger than the one of bikes ($\epsilon_{Beijing}^{bike}$=0.20), which indicates that users in Shanghai explore more unique locations than bikes over a certain period of time. While in Beijing (D2), though we still observe a sublinear scaling relation for both riders and bikes, the scaling exponents manifest an opposite pattern, where bikes are of a relatively larger exponent ($\epsilon_{Shanghai}^{bike}$=0.46) than users ($\epsilon_{Shanghai}^{user}$=0.27) (see Fig. \ref{fig:mobility-individual}d). Such discovery is consistent with the impression from Fig.~\ref{fig:basics_trajectories}f-m. 
In other datasets (D3-D8), the user-ID is not provided, only the patterns of bikes can be analysed (see Supplementary Fig. 4 and discussions in Supplementary Note 1). The exponent of bikes varies from 0.39 up to 0.73 across cities, most of which are much higher than the case of Shanghai and Beijing (see Supplementary Fig. 4a,g,m).  

As for the distributions on visitation frequency of locations $f_k$ (i.e., the revisitation dynamics) for users and bikes, both of which manifest a power-law $f_L\propto L^{-\xi}$, where $L$ is the rank of ever visited locations (the most visited one ranks first) \cite{song2010modelling}.  
It is worth noting that the scaling exponents of bikes and users are different and exhibiting opposite patterns across Shanghai (see Fig. \ref{fig:mobility-individual}b) and Beijing (see Fig. \ref{fig:mobility-individual}e). 
In Beijing (D2), riders are tending to revisit former favored locations, which is signified by a deeper slope -0.97 (against -0.69 for bikes); while in Shanghai (D1), the slopes are -0.57 and -1.14 for users and bikes, respectively, yielding an opposite pattern. The results shown in Fig.~\ref{fig:mobility-individual}a,b,d,e are qualitatively consistent with the trajectories of those most active users and bikes in Fig.~\ref{fig:basics_trajectories}f-m. 
The revisitation tendency of bikes (the slope varies between -0.72 to -1.24) in other cities (D3-D8) are also different from Beijing and Shanghai (see Supplementary Fig. 4b,h,n), and the large variation of scaling exponents cannot be explained by previous theories.

Furthermore, we also investigate 
the distribution of the longest trip for both riders and bikes. In Shanghai, it is relatively uniform for users but more concentrated among first a few trips for bikes (see Fig. \ref{fig:mobility-individual}c). 
This indicates that even after having a large number of riding trips, the next travel might still highly probably be the longest travel of a user in Shanghai; while for bikes, such probability decreases much faster. This might indicate a usage bias of users towards newer bikes.    
In Beijing, however, the distribution of users is far from uniform and decreasing in a similar way as of bikes 
(see Fig. \ref{fig:mobility-individual}f). 
Surprisingly, such distribution of bikes in Nanjing (D3 and D4) is much closer to a uniform distribution, which might indicate better maintenance of bikes in Nanjing; in other cities (D3-D8), it is closer to the case of bikes in Beijing (see Supplementary Fig. 4c).  
Figuring out the underlying mechanism behind such a phenomenon can be very important in operation optimizing of bike-sharing systems. It is thus of great significance to prolong the service life and reduce the operating cost of a bicycle if it is kept in a state that looks new and functions well. 

\subsubsection*{Population level}
\begin{figure}[ht] \centering
  \includegraphics[width=\linewidth]{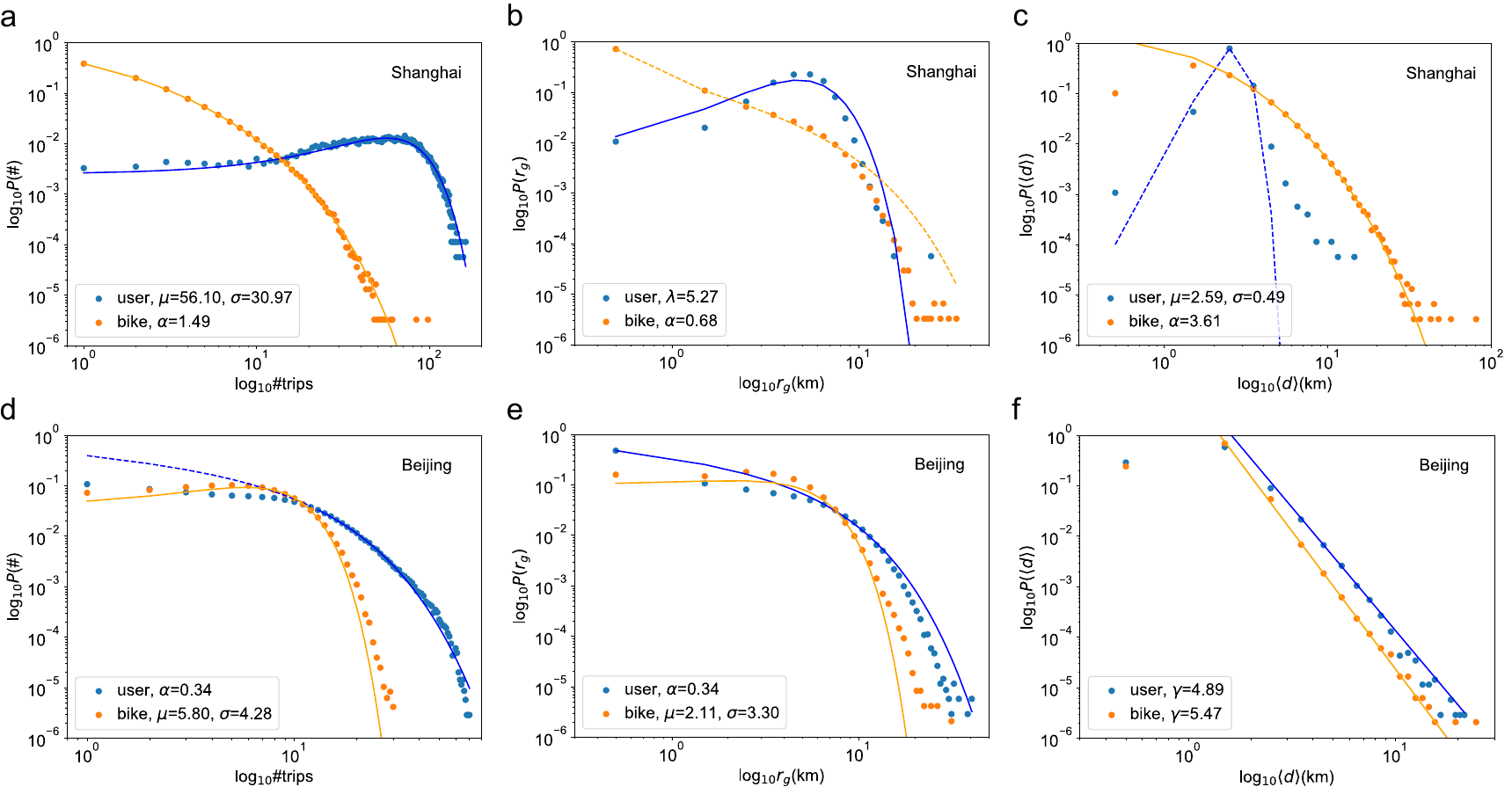} 
  \caption{Mobility patterns of riders and bikes at the population level in Shanghai (\textbf{a-c}) and Beijing (\textbf{d-f}). 
  \textbf{a, d}, The distribution of the number of trips for both users and bikes are different from each other and manifest opposite patterns across Shanghai (\textbf{a}) and Beijing (\textbf{d}). The distribution of bikes in Shanghai is well approximated by a truncated power-law, yet quite closer to a normal distribution in Beijing; as for users, they are qualitatively opposite from the patterns of bikes across cities: the users in Shanghai follow a normal distribution, while in Beijing, they follow a truncated power-law with a quite small exponent. 
  \textbf{b, e}, the distribution of gyration of users and bikes, which are different and in opposite in patterns across cities. The gyration distribution of users in Shanghai is closer to a Poisson distribution, while the distribution of bikes is closer to a truncated power-law: $P(r_g)\propto(r_g+r_{g_0})^{-\alpha} e^{-r_g/\kappa}$, where the scaling exponent $\alpha^{bike}_{Shanghai}=0.68$, the exponential cut-off $\kappa=4.91$ km, and $r_{g_0}=0.41$ km is a constant that accounts for the saturation effect in a power-law distribution.  (\textbf{b}); However, in Beijing (\textbf{e}), the distribution of users is closer to a truncated power-law with a small scaling exponent $\alpha^{user}_{Beijing}=0.34$, $\kappa=3.84$ km, $r_{g_0}=0$, and the distribution of bikes is closer to a normal distribution.  
  \textbf{c, f}, the distribution of average travel distance of users and bikes, both of which exhibit a power-law in Beijing (\textbf{f}), while the case in Shanghai (\textbf{c}) is more complex, where the distribution of bikes can be approximated by a truncated power-law, the distribution of users is closer to a normal distribution but with a not too short tail. The corresponding distributions of bikes in other cities (D3-D8) are shown in Supplementary Fig. 4, which are not the same with either city on the above indicators and manifest large variations on exponents.  
  }
 \label{fig:mobility}
\end{figure}

We further investigate several important mobility patterns at the population level. 
We find that the distribution of the number of trips for users and bikes are quite different from each other in both Beijing and Shanghai. It is a normal distribution for users and a truncated power-law for bikes in Shanghai (see Fig. \ref{fig:mobility}a). This indicates that there might be some preferential selection process on bikes. 
However, in Beijing, such patterns are almost opposite. The distribution of users is closer to a truncated power-law, 
and the distribution of bikes is quite close to a normal distribution (see Fig. \ref{fig:mobility}d). In other cities, the distributions of the number of trips of bikes is generally a normal distribution, but with Xi'an (D5) , Xiamen (D7), and Singapore (D8) evidently deviating from it at the tail part (see Supplementary Fig. 4d,j,p).  

We also analyse the distribution of radius of mobility gyration, defined as the average linear size occupied by all of his/her/its positions: $r_g = \sqrt{ \frac{1}{N}\sum_{i=1}^N(r_i-r_c)^2}$, 
where $N$ is the number of visited locations, $r_i$ denotes the (longitude, latitude) vector of i$^{th}$ location, $r_c$ is the mass centre of all locations. 
We can observe that, in Beijing (D2), the radius of gyration distribution of riders is roughly closer to a truncated power-law: $P(r_g)\propto(r_g+r_{g_0})^{-\alpha} e^{-r_g/\kappa}$, where $\alpha$ is the scaling exponent, $\kappa$ is the exponential cut-off, and $r_{g_0}$ is a constant that accounts for the saturation effect in a power-law distribution. 
This is qualitatively consistent with previous finding \cite{gonzalez2008understanding} but the scaling exponent $\alpha_{Beijing}^{user}=0.34$ is much smaller than previously reported 1.65 for all modes of travel \cite{gonzalez2008understanding}. In comparison, the distribution of bikes is relatively closer to a normal distribution with $\mu_{Beijing}^{bike}=2.11$ km in Beijing (see Fig. \ref{fig:mobility}e).  
While, in Shanghai, we did not observe similar patterns, where the distribution of riders is better fitted by a Poisson distribution with $\lambda=5.27$ km, while for bikes, it is closer to a truncated power-law (note the fitting is not good at the tail part), where $\alpha_{Shanghai}^{bike}=0.68$ (see Fig. \ref{fig:mobility}b). 
In other cities (D3-D8), the distribution of bikes can be well-fitted by a Poisson distribution except in Xi'an (D5, which deviates from it in the tail part), with $\lambda$ varies between 0.83 km to 2.87 km, and in Singapore (D8), which follows a truncated power-laws with the power exponent equal to 1.63 (see Supplementary Fig. 4e,k,q). 

In addition, the distribution of the average travel distance of riders in Shanghai (D1) is relatively closer to a normal distribution with a mean value of 2.63 km (see Fig. \ref{fig:mobility}c). 
Yet, there do exist quite a few riders (a few dozens) who deviate from the normal distribution with a relatively larger average travel distance (see Fig. \ref{fig:mobility}c). We look at these users and find that most of them have quite a small number of trips, which are long-distance travels. 
The average travel distance distribution is closer to a truncated power-law, with more bikes have a much larger average travel distance compared to a Poisson or Gaussian distribution.  
However, such patterns are not existing in Beijing, where both of them exhibit a power-law tail $P(\langle d \rangle)\propto \langle d \rangle^{-\gamma}$ with relatively similar and large scaling exponents (see Fig. \ref{fig:mobility}f). The distribution of bikes in other cities (D3-D8) also manifests a power-law tail, whose exponent $\gamma$ varies from 3.03 up to 6.27 (see Supplementary Fig. 4f). 

Since the size of users in Beijing is much larger than that in Shanghai ($\sim$ 20 times), we make a sampling analysis of users in Beijing to test whether the distributions are consistent. We find that 
the mobility patterns at both individual and population levels are quite robust (see Supplementary Fig. 5), which indicate that the biking mobility patterns might be urban characters. 
Different mobility patterns of bikes and riders in cities and large variations across cities cannot be well explained by existing theories, and such phenomena are highly probably a consequence of complex interactions between various factors, including different user-bike ratios (see Methods and Supplementary Table 1), different characteristics of public bicycling infrastructure and urban terrain, different climate and weather, promotion strategy, different level of curiosity on sharing bikes (see discussions in Supplementary Note 1), or even the attitude \cite{lizana2021analysing} towards biking. 
In the following, we develop a framework to unveil the universal scaling on choice behaviours that can well explain the inconsistent mobility patterns across cities.  

\subsection*{The emergence of scaling from the interplay between riders and sharing bikes}
Different from on-demand vehicle-sharing platforms \cite{diao2021impacts} (such as Uber, Lyft, Didi), where sharing cars are selected through optimization algorithms to reduce the waiting time of users or total travel distance \cite{santi2014quantifying,vazifeh2018addressing,molkenthin2020scaling} or to increase the revenue of the company, sharing bikes are unpowered and have to be picked up by users. So we assume that the choice behaviours are central to unlocking the puzzle presented in Figs.~\ref{fig:mobility-individual}-\ref{fig:mobility}. 
The number of available bikes, as well as the appearance (whether it looks new, clean or not) and the condition of a bike (whether it functions well, how many times it has been used), might affect users' choices on choosing which bike to ride. 

\begin{figure}[ht] \centering
  \includegraphics[width=\linewidth]{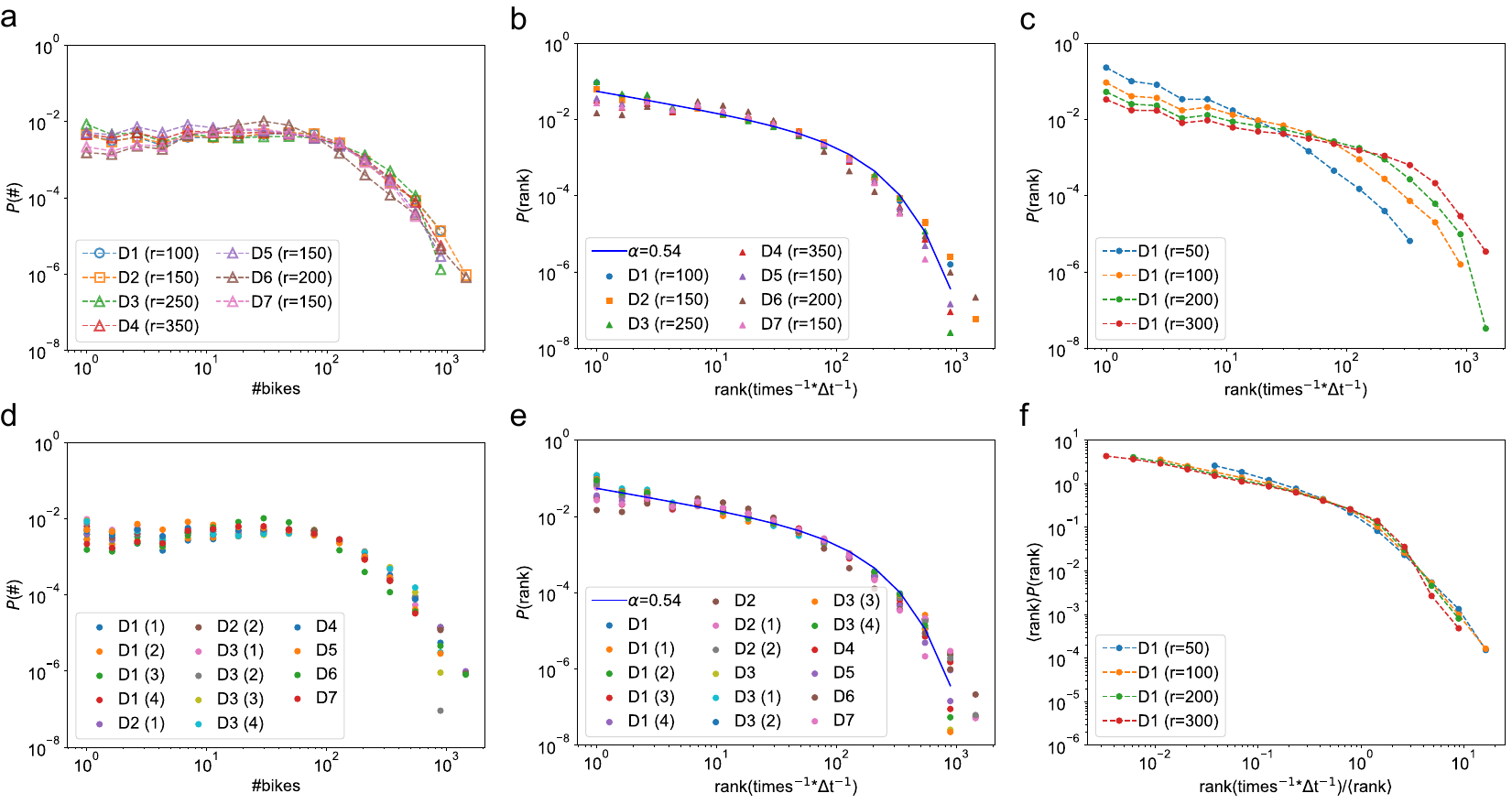}
  \caption{The scaling behaviour emerged from the interplay between riders and sharing dockless bikes. 
   \textbf{a}, The distribution of available bikes within a small region in each city, e.g., Shanghai ($r=100$m, corresponds to a grid of 200m$\times$200m), Beijing ($r=150$m), and Nanjing ($r=250$m). The curves of all these cities collapsed together, which indicates the characteristic spatial scale of cities are different, for example, $r_{Beijing}$ is roughly 1.5 times of $r_{Shanghai}$ (see Supplementary Fig. 6c for more details). When fixing $r$ across cities, the distributions never collapse together (see Supplementary Fig. 6a,b).
  \textbf{b}, the distribution of the rank value of composite indicator on the condition of the selected bike within a certain searching range $r$. Again, the curves of all these cities collapse together with different spatial scales $r$, where a scaling behaviour emerged: $P(rank)\propto rank^{-\alpha} e^{-rank/\kappa}$, with the scaling exponent $\alpha=0.54$, and exponential cut-off $\kappa=100$.  
  \textbf{c}, the distributions reported in (\textbf{b}) with varying spatial scale $r$ in Shanghai, which can well collapse together by rescaling with the average of the distribution (see \textbf{f}).
  \textbf{d}, the temporal invariance of distributions reported in (\textbf{a}) across cities.    
  \textbf{e}, the temporal invariance of distributions reported in (\textbf{b}) across cities. In (\textbf{d, e}), for all these cities, we report the distribution over the whole period, as well as in each week. The digit inside parenthesis indicates the week in the dataset, and each colour corresponds to the results obtained from a certain week. All curves collapse together with small fluctuations. 
  \textbf{f}, the rank distribution reported in (\textbf{c}) can again well collapse together by rescaling with the average of the distribution. In other cities, such collapsing across scale are also well held (see Supplementary Fig. 7). 
  }
 \label{fig:scaling}
\end{figure}

We first analyse the distribution of the number of available bikes within a certain searching range $r$ from the origin-location of each trip (i.e., a square region centred at the start location of the trip). 
Surprisingly, we find that such distributions can well collapse together across all seven cities under different spatial scales $r$ (see Fig. \ref{fig:scaling}a and Supplementary Fig. 6a). 
This indicates the characteristic spatial scales of cities are different. For example, $r_{Beijing}$ is roughly 1.5 times of $r_{Shanghai}$ (see Supplementary Fig. 6b for more details), i.e., when searching in an area with a range $r$=100 m in Shanghai, the distribution of the number of available bikes is almost the same as searching in a region with $r$=150 m in Beijing (see Fig. \ref{fig:scaling}a). 
When we fix the searching range $r$ across cities, we find that the distributions cannot collapse together (see Supplementary Fig. 6c-d). 
In addition, although literally millions 
of bikes are moving in cities every day, such distributions are also quite stable in shorter time windows. In each week, the distribution is almost the same as the ensemble one (see Fig. \ref{fig:scaling}d). 
Such collapsed distributions would be directly related to the size of bike supply as well as urban characteristics (including the size of the city, and the average size of street blocks, the width of roads, and urban terrain). The collapsed distributions might reflect some fundamental properties and intrinsic nature of the spatial distribution of sharing bikes and riding activities in cities. 

We further reveal that the distributions of the rank of the bike chosen by users are quite universal across cities and follow a scaling behaviour: $P(rank)\propto rank^{-\alpha} e^{-rank/\kappa}$, with the  scaling exponent $\alpha=0.54$, and exponential cut-off $\kappa=100$ (see Fig. \ref{fig:scaling}b). The rank of a bike in a region is calculated according to its composite condition indicator, which is considering both the condition of bikes (measured by the number of $times$ of the bike has been ridden) and the standby time $\Delta t$ since its last arrival (longer standby time $\Delta t$ usually corresponds to a higher probability of collecting dust or even bird droppings, which makes it much less attractive to riders). A larger value of composite indicator $times^{-1}\Delta t^{-1}$ corresponds to a better bike at that time and at that place (the largest value corresponds to rank 1, and so on; a smaller rank value corresponds to a better bike, and a larger rank corresponds a not-so-good one). We rank all the bikes according to such a composite condition indicator within the location whose centre is the origin of a trip. From the empirical data, we can obtain the rank value of the bike got selected by the user for each trip. 
We discover that users choose a good bike with a relatively high probability, but the scaling behaviour indicates that the probability of picking a not-so-good bike is not too small either (see Fig. \ref{fig:scaling}b). 

Such a discovery is kind of counter-intuitive, it is more natural to assume that users would prefer to ride newer and better-conditioned bikes, 
but the scaling behaviour might partially reflect limited ability on judging the condition of a bike, or, more essentially, is a result of complex interactions between various factors: whether the user is picky or not, the urgency degree of the trip, the purpose of the trip, the length of the trip, the weather condition, and the composite condition of bikes, etc. A picky user might be willing to browse more bikes to choose a satisfied one; but if the trip is urgent (e.g., to transit to a metro station to commute), then a non-broken bike might be acceptable. When the trip is long, then even a less picky user might try his/her best to find a good bike (see Supplementary Fig. 8). 
In addition, the scaling is quite robust in shorter time windows, e.g., in each week (see Fig. \ref{fig:scaling}e). 
Note that such a distribution is not derived from a static scenario, but a dynamic process. The rank value of a certain bike would change non-trivially, as new bikes would become less new and less attractive after being used or being left somewhere for a long time, then the previously lower-ranked bikes might become higher ranked ones (see Supplementary Fig. 9).  
Again, by just adjusting the searching range $r$, the rank distributions of different cities can well collapse together (see Fig. \ref{fig:scaling}b,e). 
The stability of the scaling on choice behaviours over time and across diversified cities (see Fig. \ref{fig:scaling}e) further confirms the universality. 

We then investigate the impact of searching range $r$ in each city. It mainly influences the distribution of the number of available bikes (see Fig. \ref{fig:scaling}c), and after a rescaling by the average of the distribution to take off the size effect, they can all well collapse together (see Fig. \ref{fig:scaling}f for Shanghai and Supplementary Fig. 7 for other cities).  
The collapsing of all curves indicates that regardless of the number of available choices, the scaling behaviour during the selection process is quite robust and universal across diversified cities. Our findings might be applicable in other similar selection scenarios in the era of sharing economy. 

\subsection*{Choice model that explains the mobility patterns of dockless sharing bikes}
In order to test the effectiveness of the 
discovered scaling behaviour against other human choice models, we build a multi-agent simulation system, where the origin and destination of users are streamed from the real data and users choose bikes to finish their trip based on a certain selection criterion. We implemented three different types of choice models (with six variations, see Fig.~\ref{fig:selection} and Methods and Supplementary Note 2 for more details): 

We start from the most random and unrealistic case that a user can randomly select from all available bikes in the city regardless of their spatial position as a basic null model (choice model i). Then we pose further spatial restriction that a user can only randomly choose a bike within a searching range $r$ from the origin location of the trip as an improved null model (choice model ii). Furthermore, we assume that users might tend to choose among the newest bikes. In choice model iii-iv, a user will randomly choose among the top ten bikes within the searching range. The difference between models is at the way of calculating the rank value of a bike. In choice model iii, the rank value is calculated according to the number of $times$ of the bike has been ridden (a smaller value of $times$ corresponds to a smaller rank value). Choice model iv further integrates the standby time $\Delta t$ to calculate the rank value of bikes, i.e., according to $times^{-1}\Delta t^{-1}$. Choice model v further considers the effect of varying searching range $r$. Then we come up with another choice model that a user will choose among all available bikes within the searching range according to the discovered scaling ($P(rank)\propto rank^{-\alpha} e^{-rank/\kappa}$).

In each realization in our simulation system, bikes will be chosen by users according to a certain choice model, and then the bike will be carried to the destination of the user and stayed there for a possible next trip. By tracking every move of the bike, we can eventually obtain the mobility patterns of bikes and test them against the empirical one. For each choice model, the results are obtained from averaging ten realizations. 
Compared to all other choice models, we find that the choice model that incorporates the universal scaling behaviour (choice model vi) are among the best across cities on explaining various mobility indicators (see Table \ref{tab:KS} for Beijing, Supplementary Fig. 10 and Supplementary Tables 3-4 for Shanghai and Nanjing). This confirms the effectiveness of the choice model that incorporates the discovered scaling behaviour.  

In our simulation system, when a user cannot find a bike within a certain searching range, then it will have to search for a wider area. Thus our system provides a good testing ground that if there is no bike-rebalancing process (see Supplementary Fig. 2), how long do the users need to walk to find a bike. We find that it does not take a too long distance if a user just wants to find a bike, especially in Shanghai, where the supply is relatively abundant compared to the size of users (see Supplementary Table 1); while in Beijing and Nanjing, sometimes, a much longer walking distance is needed (see Supplementary Fig. 11). But if a user wants to find a ``satisfied'' bike from a certain number of options (the number of bikes that a user need to browse is randomly selected from the distribution shown in Fig. \ref{fig:scaling}a), then the walking distance would be longer (up to a few hundred metres, see Supplementary Fig. 11). 

In addition, we notice that the Kolmogorov-Smirnov (KS) distance between the generated distribution of bikes and the empirical one on gyration is significantly larger than other indicators (see Table \ref{tab:KS} and Supplementary Tables 3-4). 
This is partially due to the daily operations we mentioned previously that a large fraction of bikes were moved since their last arrival location (see Supplementary Fig. 2). 
The distribution of bike-moving distance due to rebalancing can be roughly depicted by a truncated power-law (see Supplementary Fig. 2). In our multi-agent simulation system, we also further add such an artificial bike-moving process -- after finishing each trip, a bike will be moved if the generated random number is smaller than 2/3 (or 0.83 for Nanjing), then we extract a distance from the empirical bike-moving distance distribution (see Supplementary Fig. 2) and then a random angle from 0$^\circ$ to 360$^\circ$. Taken together, we determine the next location of the bike. With such an artificial bike-moving process, the KS distance between the generated distribution and the empirical one has dramatically dropped (see Supplementary Table 5).

\begin{table}[] \centering 
\begin{tabular}{|l|c|c|c|c|c|}
\hline
Choice models                                     & \#trips         & $\langle d \rangle$ & gyration & revisitation & longest trip \\ \hline
(i)\ \  Random (non-spatial)                        & 0.1339          & 0.0358              & 0.8331   & 0.1972       & 0.0667       \\ \hline
(ii)\  Random (spatial, $r$=100 m)                      & 0.1462          & 0.0289              & 0.7173   & 0.2872       & 0.0922       \\ \hline
(iii) $times^{-1}$ (top 10, $r$=100 m)           & 0.1948          & 0.0210              & 0.7230   & 0.2817       & 0.0954       \\ \hline
(iv)\  $times^{-1}\Delta t^{-1}$ (top 10, $r$=100 m)  & 0.1599          & 0.0284              & 0.7136   & 0.2950       & 0.1025       \\ \hline
(v)\ \  $times^{-1}\Delta t^{-1}$ (top 10, $r$=150 m) & 0.0557          & 0.0113              & \textbf{0.6789}   & \textbf{0.1949}       & 0.0264       \\ \hline
(vi)\  $times^{-1}\Delta t^{-1}$ ($r$=150 m)         & \textbf{0.0458} & \textbf{0.0094}     & 0.6949   & 0.2156       & \textbf{0.0247}       \\ \hline
\end{tabular}
\caption{The Kolmogorov-Smirnov (KS) distance between generated distributions of bikes according to various choice models (i-vi) and the empirical ones in Beijing. The generated distribution on mobility patterns of bikes is shown in Supplementary Fig. 10. Overall, the results obtained from the choice model vi that incorporates the universal scaling behaviour ($P(rank)\propto rank^{-\alpha} e^{-rank/\kappa}$) are among the best across cities on various mobility indicators (see the cases for Shanghai and Nanjing in Supplementary Tables 3 and 4, respectively).} \label{tab:KS}
\end{table}

\section*{Discussion}
In summary, newly emergent dockless bike-sharing platforms provide us a unique opportunity to study the riding behaviours in cities and the relation between users and sharing conveyances. 
Compared with docked station-based sharing bikes, dockless ones give better accessibility and more flexibility to users. 
By treating each bike as an individual entity, we first reveal that the mobility patterns of bikes and riders can be quite different from each other, especially at the population level. And mobility patterns can be inconsistent across cities. 
Reassuringly, such discrepancies can be reasonably well explained by a human choice model that incorporates a universal scaling behaviour that emerges from the interplay between travellers and sharing conveyances. Despite various diversities regarding the several cities studied, such a scaling behaviour is quite robust across cities. 
The unveiled scaling behaviour is somehow counter-intuitive: instead of picking up a good bike among the newest ones at a location, a less good bike can be chosen with a not too small probability.  
The discovery of such a simple scaling relation allows us to make better predictions on the bikes that a user chose to ride, together with a sophisticated human mobility prediction model, we can predict the whereabouts of bikes and their usage, which will be valuable to design a more efficient and user-friendly dockless bike-sharing system, and to its sustainable operation. 

We assume that dockless bike-sharing platforms aim to provide indiscriminate good-quality products and services to users, and in an ideal situation, the bikes should be utilized uniformly. Yet, this is not the case in reality (see Fig.~\ref{fig:mobility}a,d for the distribution of the number of trips of bikes) due to complex interactions between users and bikes, and the urban environment. 
We discovered that if a bike is ranked in the last quarter, it is of a higher probability of staying at that status (see Supplementary Fig. 9). 
How to ``revive'' the bike by taking advantage of the discovered scaling behaviour would be important to sustainable operations and worth future investigations. Recent advances also indicate that even a moderate increase in the financial incentives may have a disproportionately large effect on ride-sharing adoption \cite{storch2021incentive}, which might be applicable in dockless bike-sharing systems. 
For example, based on the prediction of the destination of users and the situations of bikes there, the platform can give users recommendations on bike-choosing and some sort of discount might be effective. 
In addition, the temporal evolution of interactions between individuals and evolutionary game theory \cite{liA2020evolution} can be important for better maintaining the system efficient, which is worth closer investigation in the future. 

Given the ubiquitous role that choice behaviours play in various complex systems \cite{jin2019emergence}, our results might generalize beyond the dockless bike-sharing platforms and apply to other similar selection scenarios in the era of sharing economy, which can eventually contribute towards developing a greener, healthier, and more sustainable future city.

\section*{Methods}
\subsection*{Datasets and noise filtering}
The datasets we obtained on dockless bike-sharing systems are from Mobike, DiDi Bike, and Hellobike, which are among the biggest dockless bike-sharing platforms in China. Generally, each record in the datasets has ``an order ID, a user-ID, a bike-ID, departure and arrival locations, and their corresponding timestamps''. 
The Shanghai (D1) and Beijing (D2) datasets are obtained from \textit{Mobike}, the Nanjing dataset (D3) is from \textit{DiDi Bike}. The Nanjing (D4), Chengdu (D5), and Xi'an (D6) datasets are obtained from \textit{Mobike} via online scrawler. D4-D6 are subject to stronger noises. The Xiamen dataset (D7) is aggregated from the three biggest platforms -- \textit{Mobike}, \textit{DiDi Bike}, and \textit{Hellobike}. The quality of D7 is also very high, but it is restricted to the morning rush hour from 6 a.m. to 10 a.m. in each day. The Singapore dataset (D8) is obtained from an anonymized operator there. D8 is the one shared by Ref.\cite{kondor2021estimating}, More details regarding data collection and preprocessing procedures of D8 were presented in Refs.\cite{shen2018understanding,xu2019unravelBIKE} and the supplementary materials of Ref.\cite{kondor2021estimating}.  
The Shanghai dataset (D1) and Xiamen dataset (D7) have trajectories of the trip. But user-ID is only available in D1 and D2, thus results in Figs.~\ref{fig:mobility-individual}-\ref{fig:mobility} are only available from these two datasets. In other datasets (D3-D8), we can only analyse the mobility patterns of sharing bikes and verify the findings in Fig.~\ref{fig:scaling}. In D2, the arrival time of each trip is not provided. 
Overall, D1 is of the highest quality, and thus some basic statistics in Fig. \ref{fig:basics_trajectories} are performed on this dataset. For example, in D1, there are pass-by points of the trajectory of each trip, which allows us to estimate the average riding speed. 

In D1, there are 1.02 million bill records of more than 17 thousand ``users'' (also referred as ``riders'' thereafter) and 0.3 million dockless sharing bikes over a whole month in Shanghai (from Aug. 1 to Sep. 1, 2016 -- roughly 3 months after the service of Mobike first get online); In D2, there are more than 3.2 million records, detailing the riding behaviours of 0.35 million users and 0.485 million dockless sharing bikes over two weeks (from May 10 to May 24, 2017) in Beijing; In D3, there are 1.35 million records over one month (from Oct. 1 to Oct. 31, 2019) recording the trips of 0.06 million dockless sharing bikes in Nanjing. Basics statistics of datasets (D1-D8) and more details can be found in Supplementary Tables~1-2 and Supplementary Note 1. 
  

For the raw data, we do some simple filtering: we discard the records not located within the boundary of the city and also discard the ones with a riding duration longer than one day (which might be some bikes left unlocked after their initial order) 
or less than one minute (which might be due to unsatisfied tryouts; in D1 and D3, the minimum riding duration of the raw data is one minute). 
To avoid possible biases, we do not pose any further filtering criteria. Around 200 noisy records are discarded in D1, while in D2, 0.120 million out of 3.2 million records are discarded, most of which are the ones not located in the Beijing area. In D3, there are around 0.048 million noisy records out of 1.35 million, most of which are trips less than one minute. 

\subsection*{Calculation of displacement distance.} In the Shanghai dataset (D1), the riding distance between the origin and destination of each trip is calculated from the real biking trajectory, as D1 provides the pass-by points of each trip. While in other datasets (D2-D8), no detailed riding trajectory is provided, thus the riding distance is estimated by routing distance in road networks or the straight-line Euclidean distance between the origin and destination. The routing distance is obtained from Amap API (\url{https://lbs.amap.com}, see Supplementary Note 1 for more details). 
We notice that the distribution of straight-line Euclidean distance is quite close to the distribution of routing distance, which deviates more from the real biking distance, especially for the short trips that are less than 1 km and long trips that are larger than 10 km (see Supplementary Fig. 1a).   
The distribution of Euclidean straight-line distance across cities can be reasonably well fitted by a truncated power-law $P(d)=(d+d_0)^{-\alpha} e^{-d/\kappa}$, where $\alpha$ is the scaling exponent, $d$ is the displacement distance, $d_0$ is a constant that accounts for saturation effect, and $\kappa$ is the exponential cut-off distance after which the probability would drop dramatically (see Supplementary Fig. 1b). This observation is qualitatively consistent with previous findings \cite{brockmann2006scaling,gonzalez2008understanding,alessandretti2017multi,barbosa2018human}, but the scaling exponents ($\alpha=$3.47) is much larger than the ones of all travel modes, which is usually between 1 and 2 \cite{alessandretti2017multi}. This indicates that biking trips have a much faster decay along with the increase of travel distance than other means of transportation. 

\subsection*{Choice models.} 
\begin{figure}[ht] \centering
  \includegraphics[width=\linewidth]{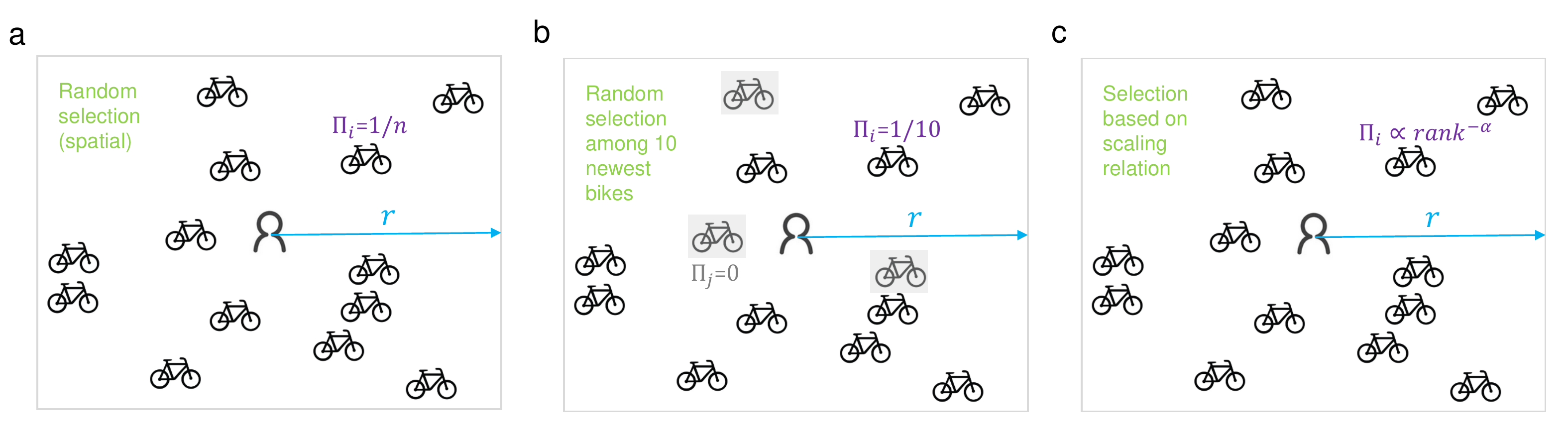}
  \caption{Choice models. 
  \textbf{a}, random selection. As for the most random and unrealistic null model (choice model i, which is also described as ``random (non-spatial)'' in Table \ref{tab:KS}), $r$ can be as large as the diametre of the city. The probability of a bike $i$ got selected is $\Pi_i=1/N$, where $N$ is the number of all available bikes in the whole city.    The spatial one (choice model ii) is selecting from the area with a certain searching range $r$. The probability of a bike $i$ got selected is $\Pi_i=1/n$.   
  \textbf{b}, choosing among the newest ones, which includes three variations. The condition of bikes can be assessed by the number of $times$ a bike has been ridden (e.g., $times^{-1}$ (top 10, $r$=100 m)), or the composite indicator that considers stand-by duration since its last arrival $\Delta t$ (e.g., $times^{-1}\Delta t^{-1}$ (top 10, $r$=100 m)), and the searching range $r$ can be varying (e.g., $times^{-1}\Delta t^{-1}$ ($r$=200 m)). The bikes in grey shading are ones that fall out top ten according to a certain criterion. Then $\Pi_i=1/10$ for those ten newest bikes, and for those that fall out the top ten, the selection probability $\Pi_j=0$. 
  \textbf{c}, choice model that incorporates the scaling behaviour: $\Pi_i\propto rank^{-\alpha}e^{-rank/\kappa}$ (choice model vi in the main text). See Methods for more details of choice models. The icons used in this figure are work of Azaze11o/Shutterstock.com. 
  }
 \label{fig:selection}
\end{figure}

In each realization in our multi-agent simulation systems, users will choose a bike based on a certain choice model. The trip will be associated with the chosen bike and the user. Then we can obtain the mobility pattern of bikes based on the simulated data and compare them with the empirical one to get the corresponding KS distance. The results shown in Supplementary Fig. 10 and Table \ref{tab:KS} are the average of ten realizations. 
we implement six variations of three different types of choice models (see Fig.~\ref{fig:selection}) as follows: 

\textbf{Random selection:}
\begin{itemize}
  \item Choice model i: Random (non-spatial). This is the most unrealistic and the most random scenario that the user in the simulation platform can randomly choose any bike in the city regardless of its spatial position. The probability of a bike $i$ got selected $\Pi_i=1/N$, where $N$ is the total number of bikes in a city. This is regarded as a basic null model.
  \item Choice model ii: Random (spatial, $r$=100 m): This mechanism is more realistic to consider the spatial constraints in reality that a user can only pick up a bike within a certain searching range $r$ (here, $r=$100 m) from the start location of his/her trip. The probability of a bike $i$ got selected $\Pi_i=1/n$, where $n$ is the total number of bikes within the searching range. This is regarded as an improved null model. 
\end{itemize}

\textbf{Choosing among the newest ones:}
\begin{itemize}
  \item Choice model iii: $times^{-1}$ (top 10, $r$=100 m).  Furthermore, we assume that the condition of a bike would also have an impact, so we calculate the number of trips (marked as $times$) that a bike has been taken as a proxy to the condition of a bike. It is natural to assume that a user is preferring to use a newer bike. A larger value of $times$ would usually mean more serious wear of the bike itself or even a higher probability of riding by a reckless user who does not care for the bike. In this choice model, a user will randomly choose among the ten newest bikes ranked by $times^{-1}$ within the searching range $r$=100 m from the origin of the trip. The probability that a bike $i$ got selected $\Pi_i=1/10$, if $i$ is among the top 10; otherwise, $\Pi_i=0$.
  
  \item Choice model iv: \textit{times}$^{-1}\Delta t^{-1}$ (top 10, $r$=100 m). Compared to choice model iii, we assume that the appearance of a bike also influences the choice of a user, thus we calculate the standby time of the bike since its last arrival (marked as $\Delta t$). Longer standby time $\Delta t$ would usually correspond to a higher probability of collecting dust or even bird droppings, which makes it much less attractive to riders. Similarly, in this model, a user would choose among the ten newest bikes ranked by $times^{-1}\Delta t^{-1}$ within the searching range $r$=100 m. The form of $\Pi_i$ is the same as choice model iii. 
  \item Choice model v: \textit{times}$^{-1}\Delta t^{-1}$ (top 10, $r$=150 m). In addition, the searching range $r$ might play a role. In this model, a user would randomly choose among the ten newest bikes ranked by $times^{-1}\Delta t^{-1}$ within a larger searching range. The form of $\Pi_i$ is the same as choice model iii. 
\end{itemize}


\textbf{Choice model that incorporates the discovered scaling behaviour:}
\begin{itemize}
  \item Choice model vi: $times^{-1}\Delta t^{-1}$ ($r$=150 m). A user will choose among all available bikes within the searching range $r$=200 m, and the probability that a bike got selected 
  $\Pi_i \propto rank_i^{-\alpha} e^{-rank_i/\kappa}$, with the scaling exponent $\alpha=0.54$, exponential cut-off $\kappa=100$. For Beijing, $r$=150 m, and for other cities, the value of $r$ is in accordance with the values presented in Fig. \ref{fig:scaling}a,b.
\end{itemize}

More details of these choice models can be found in Supplementary Note 2. 

\section*{Data availability}
The original datasets necessary to reproduce the results in the manuscript are available from the website of Soda competition (\url{http://shanghai.sodachallenges.com/data.html?lang=en}) for the Shanghai dataset (D1), and Mobike Cup competition (\url{https://biendata.com/competition/mobike/data/}) for the Beijing dataset (D2). 
The Nanjing dataset (D3) is not publicly available due to the commercially sensitive information contained and the non-disclosure agreement with the DiDi Bike, but the data needed to reproduce this work is available from the corresponding author on reasonable requests. Other datasets (D4-D6) are available at \url{https://github.com/UrbanNet-Lab/scaling_in_dockless_sharing_bikes}. 
The Xiamen dataset (D7) is available from the website of Digital China Innovation Contest (\url{https://data.xm.gov.cn/contest-series/digit-china-2021/#/3/competition_data}). 
The Singapore dataset (D8) is available from Ref.\cite{kondor2021estimating} and at \url{https://github.com/dkondor/bikesharing_data}. 

\section*{Code availability}
The code used in the manuscript is available at \url{https://github.com/UrbanNet-Lab/scaling_in_dockless_sharing_bikes}.

\bibliography{mobike}

\section*{Acknowledgements}
We acknowledge financial supports from the National Natural Science Foundation of China (Grant No. 61903020), Fundamental Research Funds for the Central Universities (Grant No. buctrc201825). 
L.L. acknowledges financial supports from the National Natural Science Foundation of China (Grants Nos. 61673150, 11622538), the Science Strength Promotion Program of the University of Electronic Science and Technology of China (Grant No. Y030190261010020). 
R.L. acknowledges Dr. Qing Yao from Beijing Normal University, Dr. Gezhi Xiu and Dr. Jianying Wang from Peking University for helpful discussions. A.L. acknowledges technical help from Ms. Shuai Gao from UrbanNet Lab.

\section*{Author contributions statement}
R.L. conceived and designed the research, J.F., L.L., and H.E.S. refined the research, A.L. and F.S. analysed the empirical data, R.L., A.L., J.F., and L.L. analysed the results. R.L. was the lead writer of the manuscript. All authors reviewed the manuscript. 

\section*{Competing interests}
The authors declare no competing interests.

\clearpage

\section*{Supplementary Information for \\Emergence of Scaling in Dockless Bike-sharing Systems}

\renewcommand\figurename{Supplementary Figure}
\renewcommand\tablename{Supplementary Table}

\setcounter{figure}{0}   

\begin{figure}[htbp!] \centering
  \includegraphics[width=0.8\linewidth]{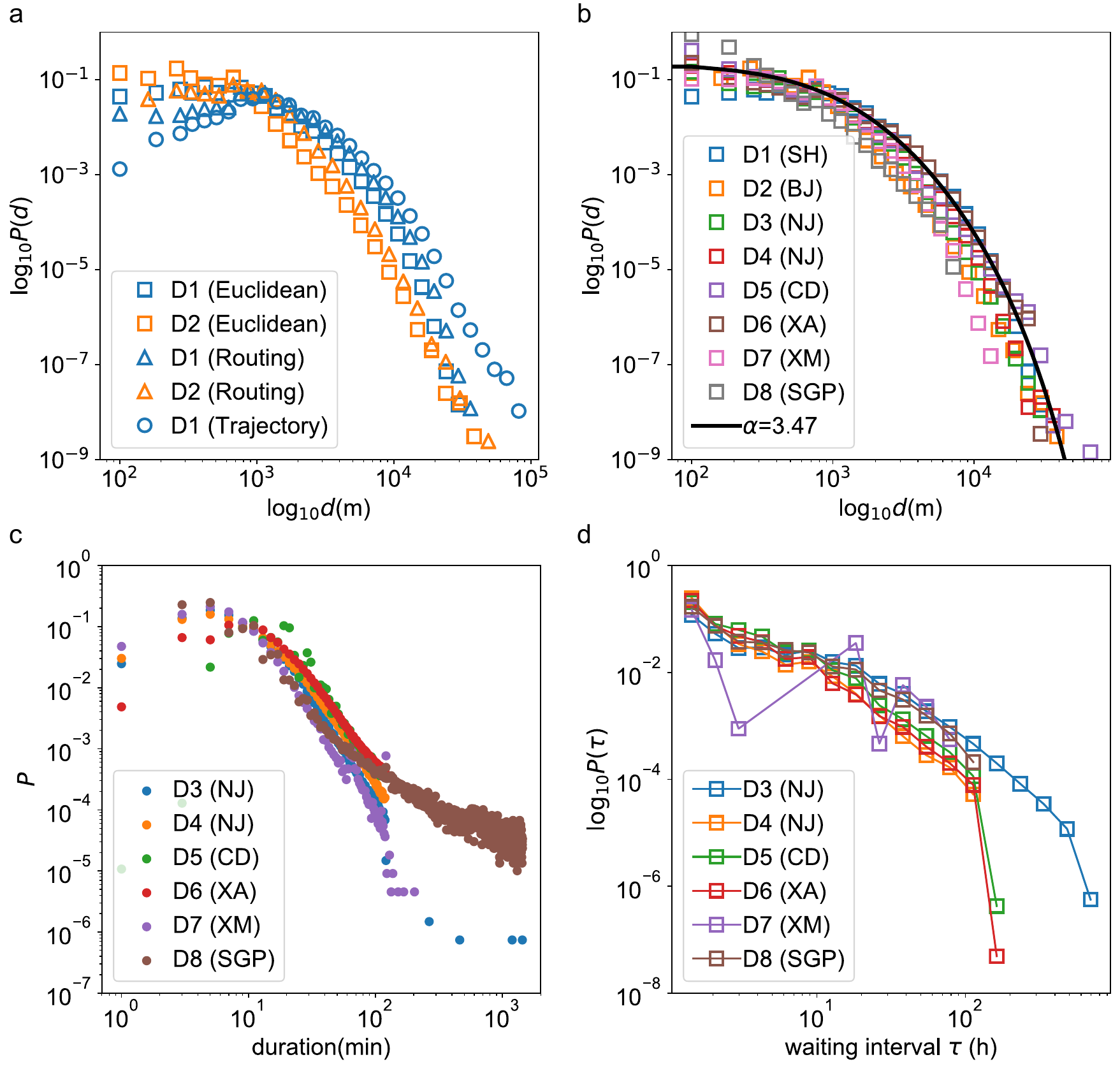}
  \caption{The distribution of displacement distance (\textbf{a,b}), riding duration (\textbf{c}), and inter-trip waiting time (\textbf{d}). 
  \textbf{a}, the distribution of displacement distance that is calculated based on different criteria. For Shanghai (D1), we report the real travel distance ({\scriptsize $\Circle$}), the shortest routing distance ({\scriptsize $\triangle$}), and the straight-line Euclidean distance between the origin and destination of the trip ({\scriptsize $\square$}). 
  The real travel distance is calculated from the biking trajectory of the trip, and the routing distance is obtained from Amap API (\url{https://lbs.amap.com}, see Supplementary Note 1 for more details).  
  For Beijing (D2), as there is no trajectory provided (see Supplementary Table 2 for more details), we only compare the routing distance ({\scriptsize $\triangle$}) and Euclidean distance ({\scriptsize $\square$}). We can observe that the real travel distance is significantly longer than the Euclidean one in Shanghai, while the routing distance is relatively closer to the Euclidean one in both cities. 
  \textbf{b}, the distribution of straight-line displacement distance across cities (D1-D8) can be reasonably well fitted by a truncated power-law $P(d)=(d+d_0)^{-\alpha} e^{-d/\kappa}$ across cities with $\alpha=3.47,\ d_0=1884\ \mathrm{m},\ \kappa=5347\ \mathrm{m}$. 
  \textbf{c}, the distribution of riding duration of trips of D3-D8 all exhibits a power-law tail. We suspect that most platforms have already performed some filterings before sharing the data, as there is a sudden drop of trips after 120 minutes for both D3 and D7. Since D2 does not have the arrival time (see Supplementary Table 2), its riding duration is not reported here.  
  \textbf{d}, the distributions of waiting interval between two consecutive trips of bikes across cities are closer to a power-law. As D7 is incomplete, which only has data from 6 a.m. to 10 a.m. in each day, thus its corresponding distribution is zigzagged and deviates from others. 
  D3-D6 and D8 exhibit a similar power-law tail, yet differ from the distributions of bikes in D1 and D2 (see Fig. 1e in the main text). 
  }
 \label{fig:waitingInterval_duration}
\end{figure}

\newpage
\begin{figure}[htbp!] \centering
  \includegraphics[width=0.5\linewidth]{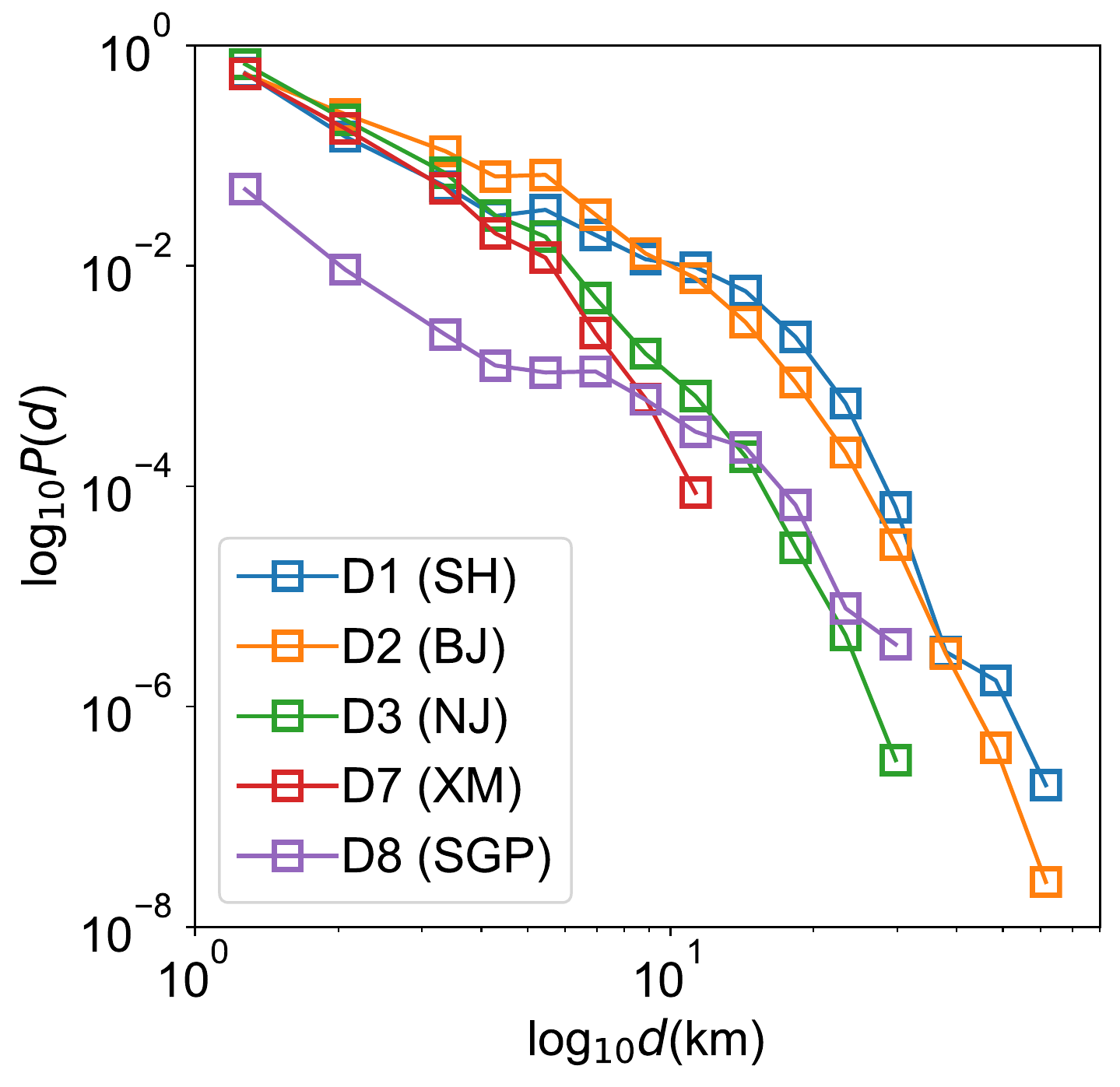}
  \caption{The distribution of moving distance of bikes due to daily rebalancing operations and maintenance in Shanghai (D1), Beijing (D2), Nanjing (D3), Xiamen (D7), and Singapore (D8).    
  The moving of a bike is identified as a different start location from the last arrival location, e.g., a bike arrived a location A in the last trip, but it might later start a new trip from a different location B. This is mainly due to the daily operations of the company to rebalance bikes between locations. There are roughly 2/3 of bikes in both D1 and D2 have been moved from their last arrival locations, while in D3, D7, and D8, the fraction reaches 83\%, 55\%, and 34\%, respectively. As D4-D6 are crawled from the platform, the bike-moving due to rebalancing cannot be identified (see Supplementary Note 1 for more details).} 
 \label{fig:moving_distance}
\end{figure} 

\begin{figure}[htbp!] \centering
  \includegraphics[width=0.5\linewidth]{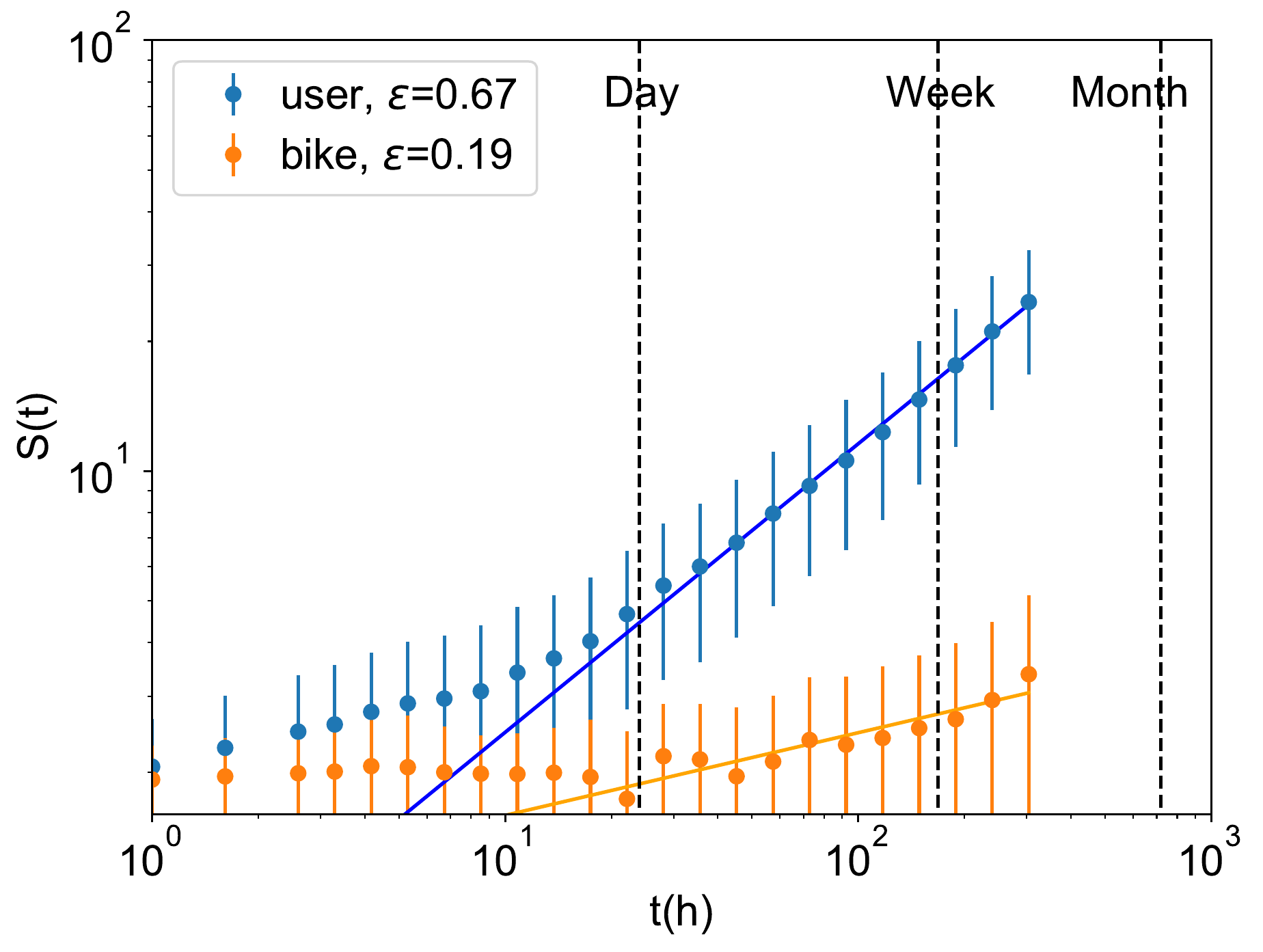}
  \caption{The exploring dynamics of bikes and riders in Shanghai (D1) in the first two weeks. The patterns and exponents are almost identical to the case of the whole month reported in Fig. 2a in the main text. The scaling exponent of users is larger than the one of bikes, which indicates that users in Shanghai tend to explore more unique locations than bikes.}
 \label{fig:exploration}
\end{figure} 

\newpage
\begin{figure}[ht] \centering
  \includegraphics[width=0.75\linewidth]{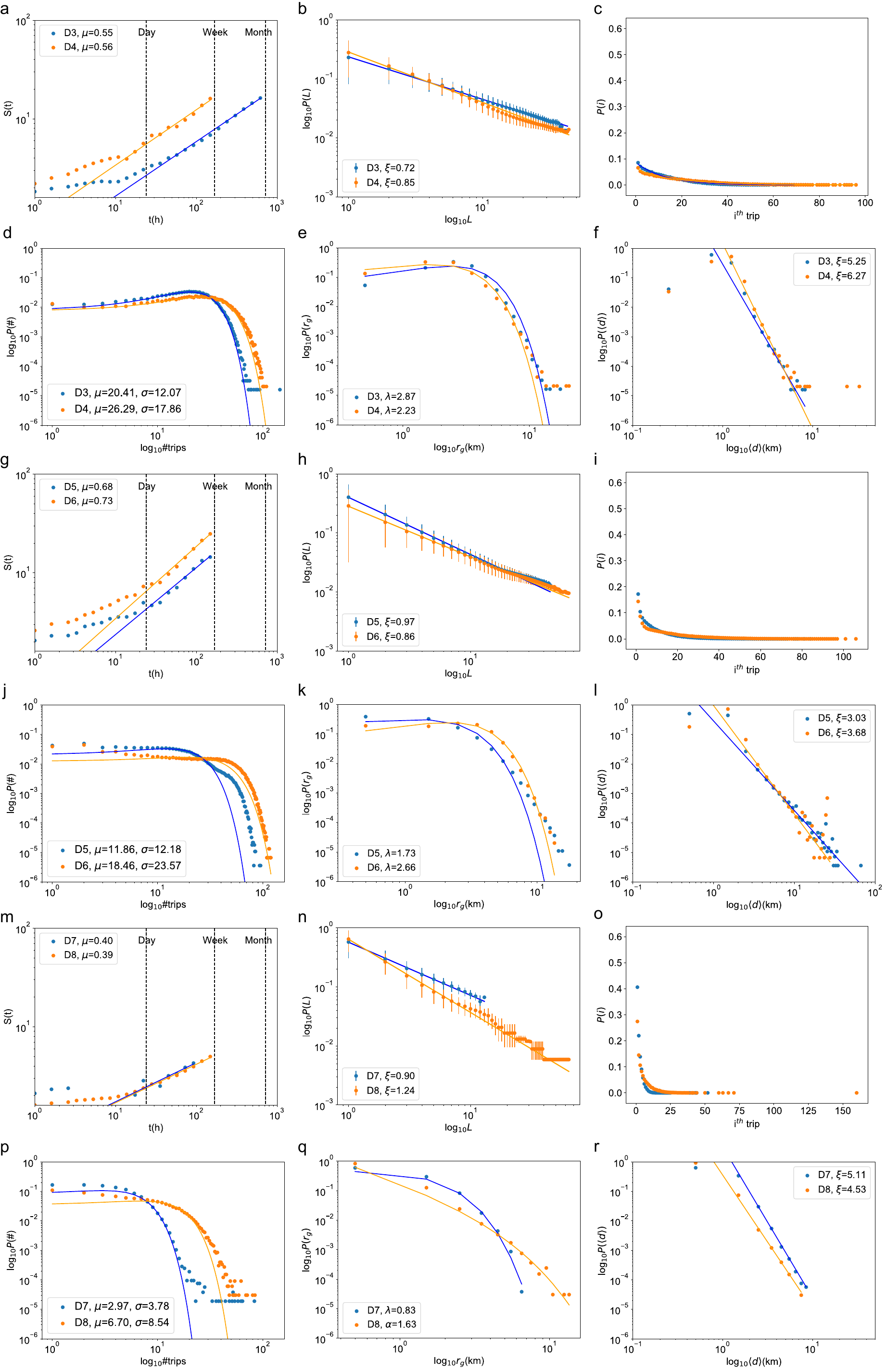}
  \caption{Mobility patterns of bikes at both individual level (\textbf{a-c, g-i, m-o}) and population level (\textbf{d-f, j-l, p-r}) in (\textbf{a-f}) Nanjing (D3, D4), (\textbf{g-l}) Xi'an (D5), Chengdu (D6), (\textbf{m-r}) Xiamen (D7), and Singapore (D8).  
  \textbf{a, g, m}, the exploration of new locations of bikes along with the increase of observation duration. The scaling exponent varies between 0.39 and 0.73. 
  \textbf{b, h, n}, the revisitation dynamics on locations of bikes. The slope is larger than the cases in both Beijing and Shanghai in Fig. 2b,e in the main text. The scaling exponent varies between 0.72 and 1.24. 
  \textbf{c, i, o}, the distribution of the longest trip of bikes. The longest trip of a bike is quite uniform in Nanjing (D3, D4), which is quite different from other cities (see Fig. 2c,f in the main text and Supplementary Fig. \ref{fig:mobilityPatternsBike}i,o). 
  \textbf{d, j, p}, The distribution of trips of bikes, which is well approximated by a Normal distribution. It is worth noting that D5, D7, and D8 deviate from it at the tail part.  
  \textbf{e, k, q}, the distribution of gyration of bikes, which is quite close to a Poisson distribution with $\lambda$ varies between 0.83 and 2.87 for D3-D7, and close to a truncated power-law with an exponent equals 1.63 for D8.  
  \textbf{f, l, r}, the distributions of average travel distance of bikes, which exhibit a power-law that is similar to the case of Beijing in Fig. 3f in the main text. The exponent varies between 3.03 and 6.27.}
 \label{fig:mobilityPatternsBike}
\end{figure} 


\clearpage

\begin{figure}[ht] \centering
  \includegraphics[width=\linewidth]{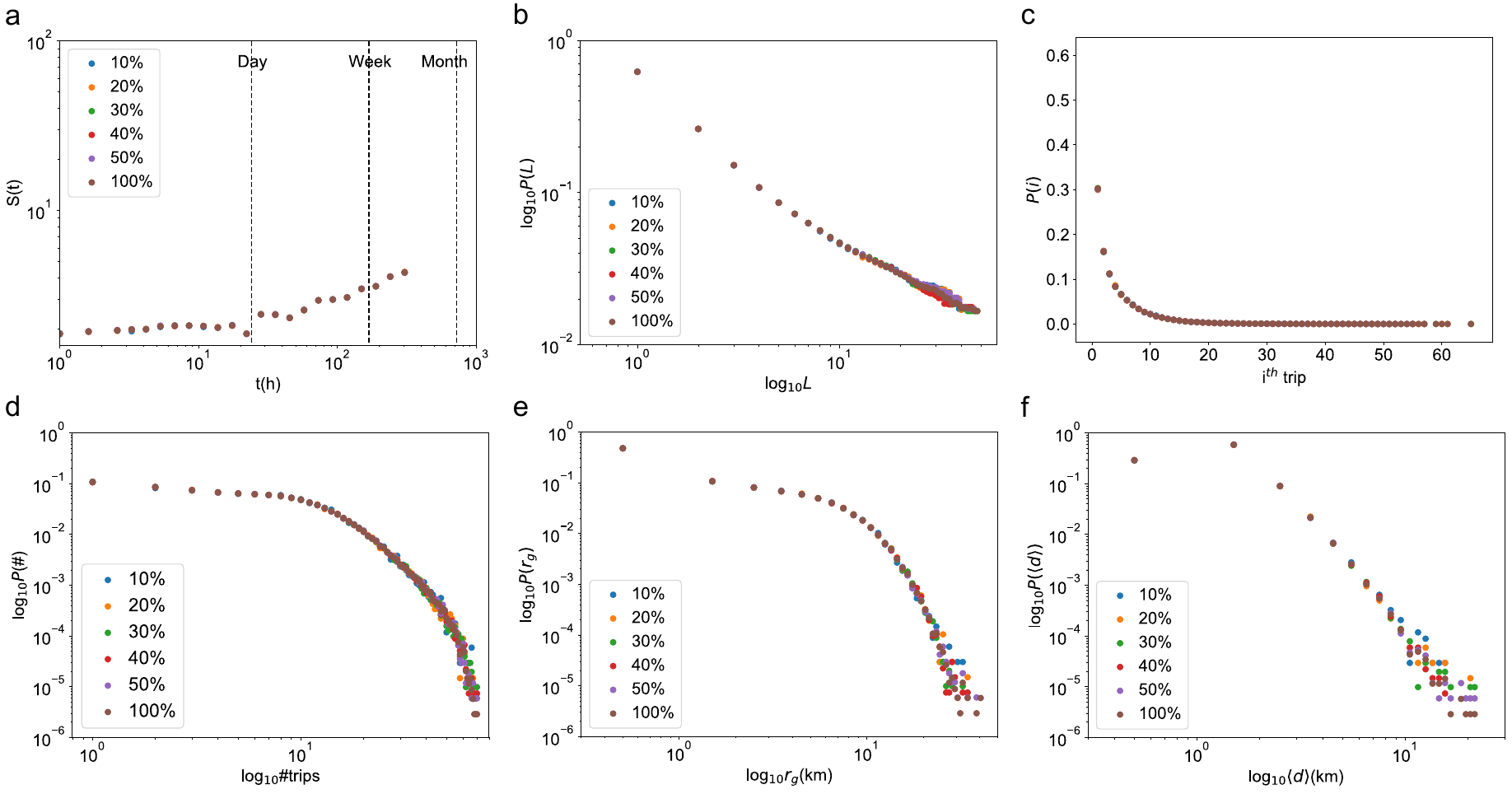} 
  \caption{The mobility patterns of sampled users of D2 with a sample size from 5\% (equal to the size of users of D1) up to 50\%. As the mobility patterns of D1 and D2 are quite different from each other (see Figs. 2-3 in the main text), we wonder that if it is induced by different size of users (as there were only 0.017 million users in D1, but 0.350 million in D2, which is of roughly 20 times difference). The results presented here show that the size of users cannot explain the difference across cities, and the mobility patterns of users are quite robust over different sampling rate.}
 \label{fig:sampling}
\end{figure}


\clearpage
\begin{figure}[htb] \centering
  \includegraphics[width=0.8\linewidth]{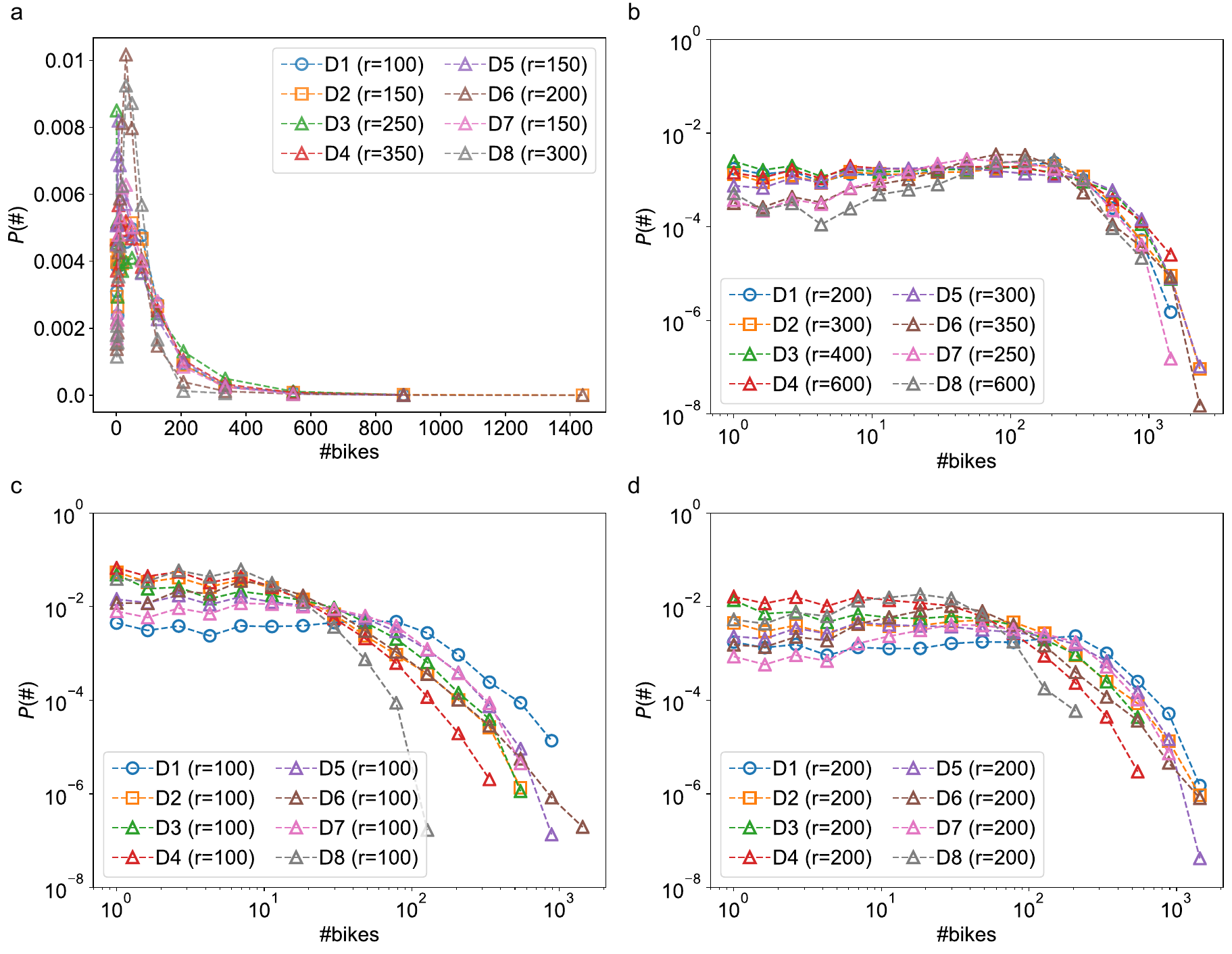}
  \caption{The distribution of the number of bikes in locations with  (\textbf{a, b}) varying and (\textbf{c, d}) fixed searching range $r$ across cities (D1-D8).  
  \textbf{a}, the same distribution as of Fig. 4a in the main text in linear axis. 
  \textbf{b}, the similar distribution as of Fig. 4a in the main text with larger searching ranges $r$ across cities. Again, under different spatial scales, the distributions well collapse together (with slight deviations in the head part of D6-D8). And the multiplicators are quite comparable with the ones in Fig. 4a in the main text. For example, $r_{Shanghai}$=200 m, $r_{Beijing}$=300 m, which is still 1.5 times of Shanghai.  
  \textbf{c}, when the searching range $r$ is fixed as 100 m across cities, the distributions of the number of bikes never collapse together.
  \textbf{d}, when the searching range $r$ is fixed as 200 m across cities, the distributions of the number of bikes also never collapse together.
  }
 \label{fig:numberDistribution}
\end{figure} 

\clearpage

\begin{figure}[!htb] \centering
  \includegraphics[width=0.8\linewidth]{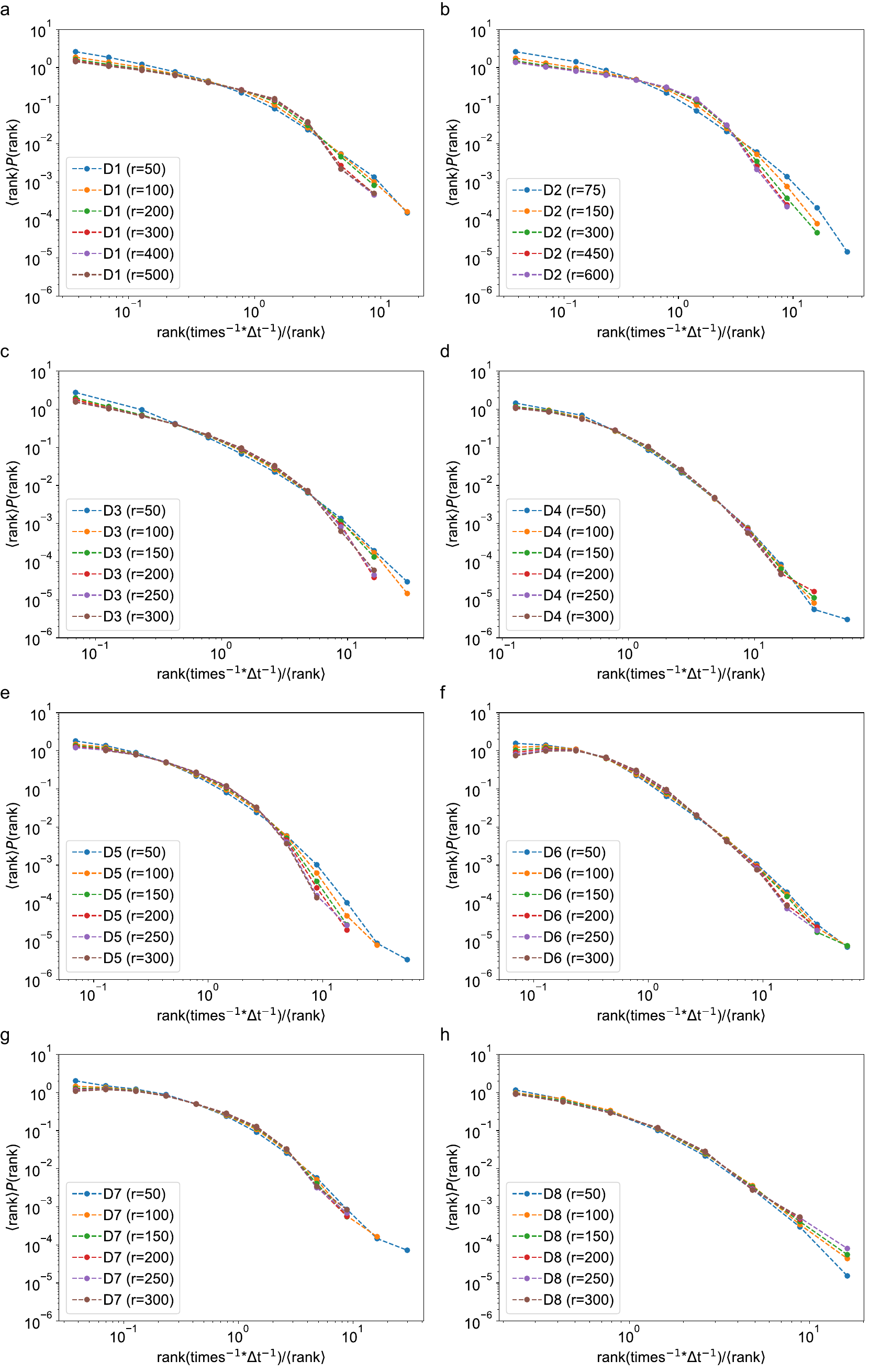}
  \caption{The rescaled rank distributions of the selected bikes in locations with varying searching range $r$ across cities (D1-D8).   
  \textbf{a}, Shanghai (D1). 
  \textbf{b}, Beijing (D2). 
  \textbf{c}, Nanjing (D3).
  \textbf{d}, Nanjing (D4). 
  \textbf{e}, Chengdu (D5).
  \textbf{f}, Xi'an (D6). 
  \textbf{g}, Xiamen (D7). 
  \textbf{h}, Singapore (D8).
  The distributions in each city all well collapse together after rescaled by its average. This further confirms the universality of the scaling behaviours that emerges from the interplay between users and sharing bikes.}
 \label{fig:rescaledRankDistribution}
\end{figure}

\begin{figure}[!htbp] \centering
  \includegraphics[width=0.6\linewidth]{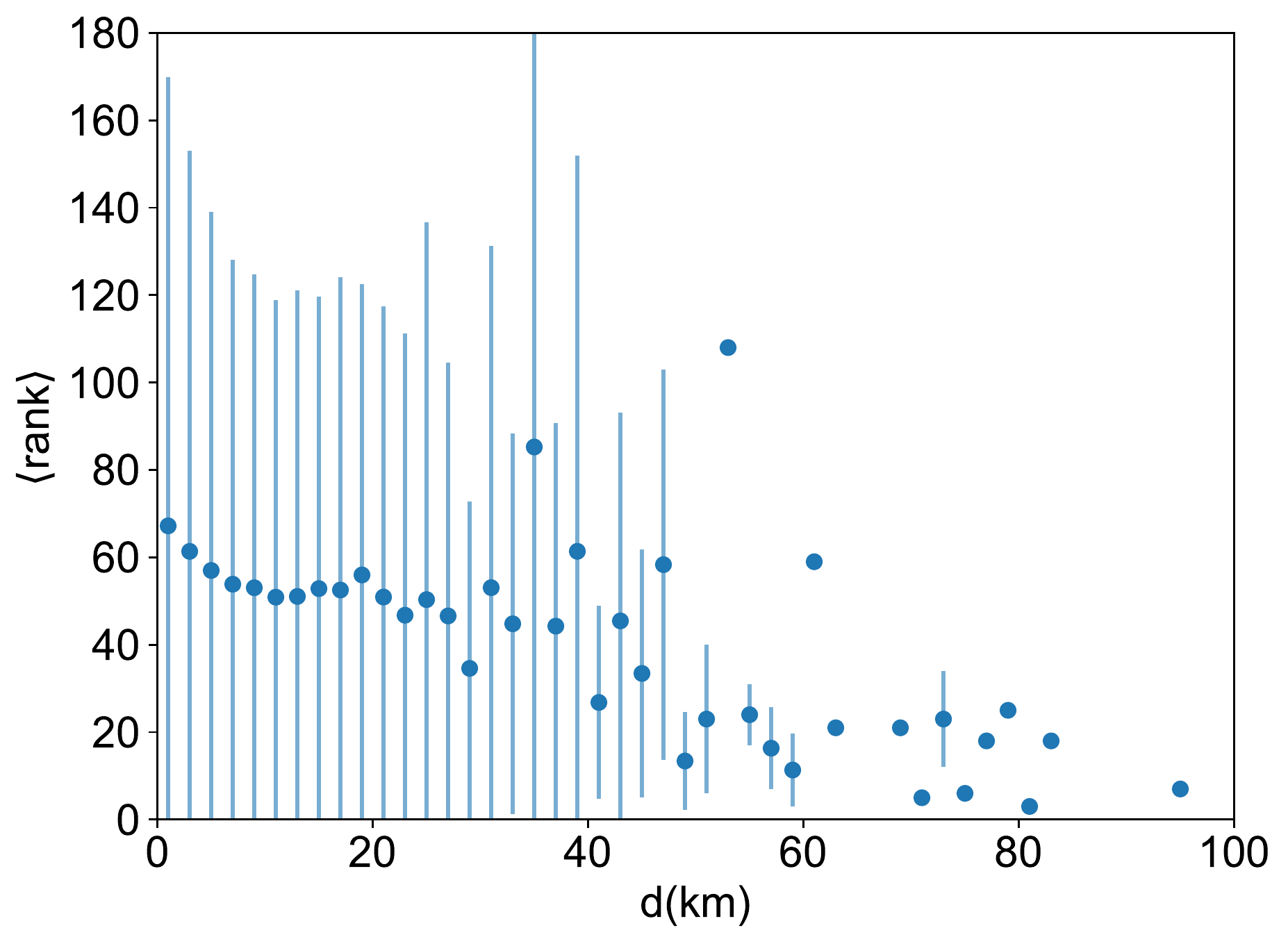}
  \caption{The average rank of bikes got selected for trips with increasing trip length in Shanghai (D1). It is clear that for longer distance trips, the average rank value of bikes got selected is smaller (i.e., the composite condition of the bike is better). Error bars mean standard deviation.} 
 \label{fig:averank-tripdis}
\end{figure}

\begin{figure}[ht] \centering
  \includegraphics[width=\linewidth]{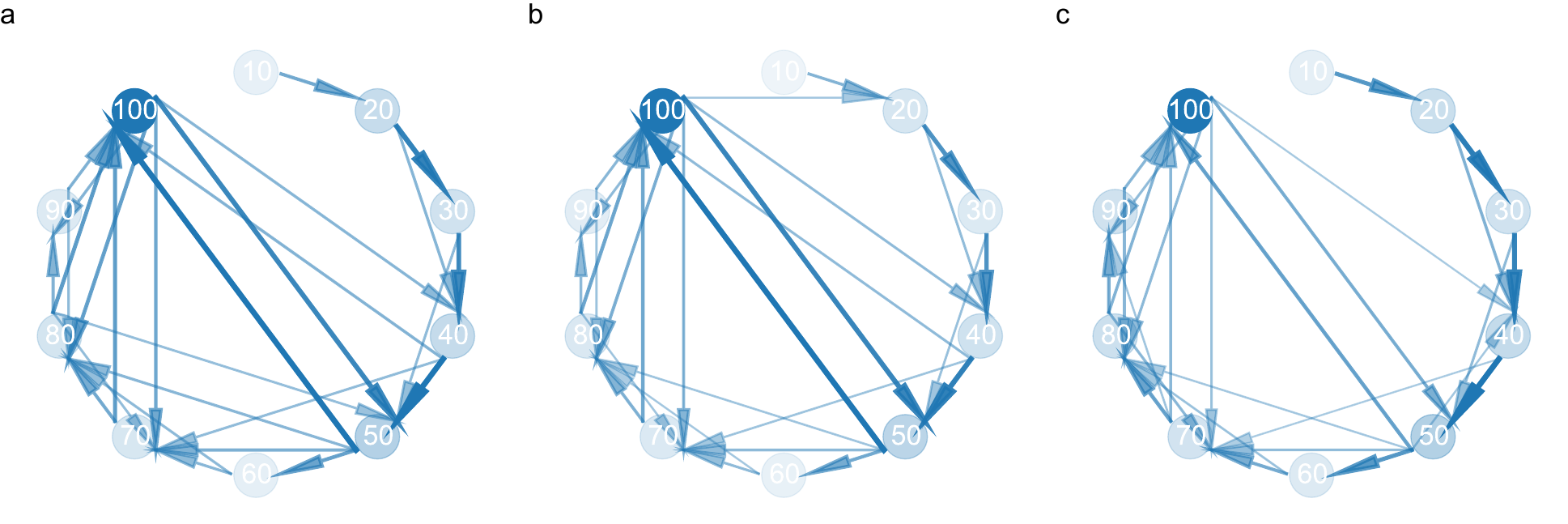}
  \caption{The dynamics of the ranking of bikes. At each hour, we calculate the rank of the bike within the searching range that is consistent with Fig. 4a in the main text for each city. The rank value of each bike is categorised into ten levels: first ten percent (indicated by the ``10'' in the node, i.e., the newest ones), up to 90-100 percent (``100'', the most unwanted ones). We observe a clear descending trend, but much less a ``reviving'' ascending trend. And the patterns are similar across cities. For clarity, we only show the top 30\% of edges with the highest volume, and the color of the node denotes the self-loop volume. The results are obtained from the first day of the data as an example. \textbf{a}, Shanghai (D1); \textbf{b}, Beijing (D2); \textbf{c}, Nanjing (D3). }
 \label{fig:bikeDynamics}
\end{figure}

\clearpage
\begin{figure}[ht] \centering
  \includegraphics[width=0.8\linewidth]{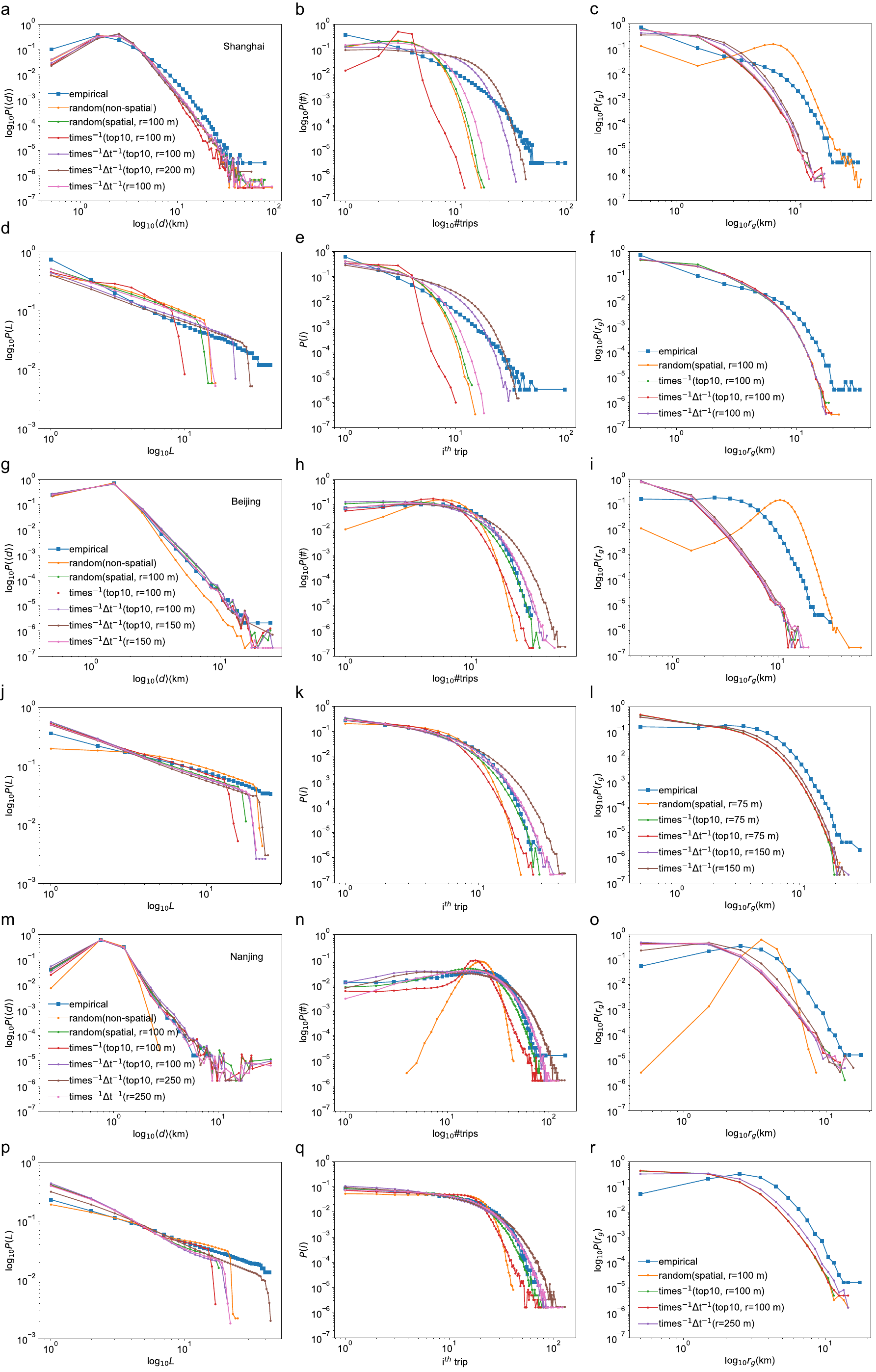}
  \caption{The empirical and generated mobility patterns of bikes in our multi-agent simulation system at both population level (\textbf{a-c, g-i, m-o}) and individual level (\textbf{d-f, j-l, p-r}) in Shanghai (\textbf{a-e}), Beijing (\textbf{g-l}), and Nanjing (\textbf{m-r}). They are generated by a variety of choice models based on real spatio-temporal biking demands. This figure is complementary to results presented in the Table 1 \textbf{(g-k)} in the main text and Supplementary Tables 3-4 \textbf{(a-e, m-q)} and Supplementary Table 5 \textbf{(f, l, r)}. 
  Solid lines in the figure are guidance to eyes. 
  } 
 \label{fig:predictedBikeMobility}
\end{figure}

\clearpage
\begin{figure}[ht] \centering
  \includegraphics[width=\linewidth]{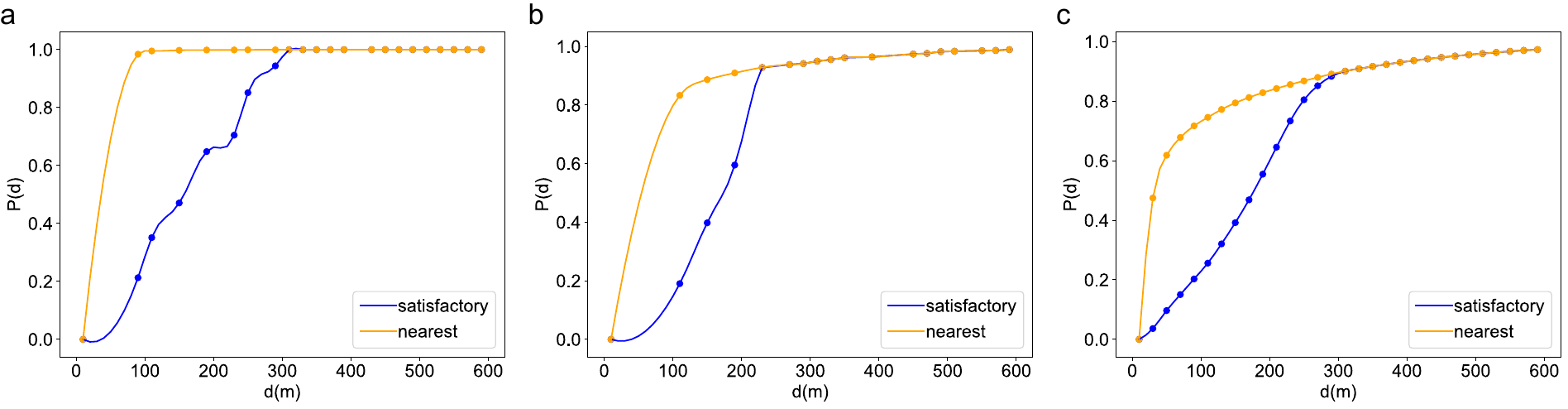}
  \caption{Without rebalancing bikes, the distance that a user needs to walk to find a bike under two different criteria: one is finding the nearest one (orange lines), the other is finding a satisfied one from a certain size of options (blue lines). As for the latter criterion, we first randomly withdraw a number from the distribution shown in Fig. 4a in the main text, as we assume that the user will only make a choice after browsing enough bikes. The simulation results for (\textbf{a}) Shanghai, (\textbf{b}) Beijing, and (\textbf{c}) Nanjing are reported. We can observe that the user does not need to walk a too long distance to find a bike, especially in Shanghai; while if the user has to browse a certain number of bikes to make the choice, then the walking distance needed sometimes can be a few hundred metres. 
 }
 \label{fig:additionalDistance}
\end{figure}


\clearpage


\begin{table}[] \centering
\begin{tabular}{|l|l|l|l|l|l|}
\hline
\multicolumn{1}{|c|}{City} & \multicolumn{1}{c|}{Platform} & \multicolumn{1}{c|}{Rec.} & \multicolumn{1}{c|}{Period(MM/DD)} & \multicolumn{1}{c|}{Users} & \multicolumn{1}{c|}{Bikes} \\ \hline
Shanghai (D1)                    & Mobike                        & 1.02                               & 08/01-09/01, 2016        & 0.017                            & 0.300                             \\ \hline
Beijing (D2)                   & Mobike                        & 3.20                               & 05/10-05/24, 2017        & 0.350                             & 0.485                            \\ \hline
Nanjing (D3)                    & DiDi Bike                     & 1.35                               & 10/01-10/31, 2019        & NA                                & 0.060                             \\ \hline
Nanjing (D4)                    & Mobike                        & 1.44                               & 03/20-03/26, 2017        & NA                                & 0.048                            \\ \hline
Chengdu (D5) & Mobike & 4.40 & 09/03-09/09, 2018 & NA & 0.275\\ \hline
Xi'an (D6) & Mobike & 4.09 & 09/03-09/09, 2018 & NA & 0.150 \\ \hline
Xiamen (D7) & aggregated & 0.22 & 12/21-12/25, 2020 & NA & 0.053 \\ \hline
Singapore (D8) & NA & 0.30 & 09/11-09/17, 2017 & NA & 0.033 
\\ \hline
\end{tabular}
\caption{Basic statistics of datasets. The D7 is aggregated from the three biggest platforms in Xiamen -- Mobike, DiDi Bike, and Hellobike. Note that D7 only has data during the morning rush hours from 6:00 a.m. to right before 10:00 a.m., not whole days as in D1-D6. NA stands for Not Available. The numbers of records, users, and bikes are in Million.} \label{tab.statistics}
\end{table}

\begin{table}[] \centering
\begin{tabular}{|c|c|c|c|c|c|c|c|c|}
\hline
Dataset           & D1                       & D2                       & D3                       & D4                       & D5                       & D6    & D7       &D8             \\ \hline
UserID            & {\color[HTML]{32CB00} Y} & {\color[HTML]{32CB00} Y} & {\color[HTML]{FD6864} N} & {\color[HTML]{FD6864} N} & {\color[HTML]{FD6864} N} & {\color[HTML]{FD6864} N} & {\color[HTML]{FD6864} N} & {\color[HTML]{FD6864} N} \\ \hline
BikeID            & {\color[HTML]{32CB00} Y} & {\color[HTML]{32CB00} Y} & {\color[HTML]{32CB00} Y} & {\color[HTML]{32CB00} Y} & {\color[HTML]{32CB00} Y} & {\color[HTML]{32CB00} Y} & {\color[HTML]{32CB00} Y} & {\color[HTML]{32CB00} Y} \\ \hline
DepartureTime     & {\color[HTML]{32CB00} Y} & {\color[HTML]{32CB00} Y} & {\color[HTML]{32CB00} Y} & A                        & A                        & A   & {\color[HTML]{32CB00} Y}              & {\color[HTML]{32CB00} Y}                     \\ \hline
DepartureLocation & {\color[HTML]{32CB00} Y} & {\color[HTML]{32CB00} Y} & {\color[HTML]{32CB00} Y} & A                        & A                        & A   & {\color[HTML]{32CB00} Y}              & {\color[HTML]{32CB00} Y}                     \\ \hline
ArrivalTime       & {\color[HTML]{32CB00} Y} & {\color[HTML]{FD6864} N} & {\color[HTML]{32CB00} Y} & A                        & A                        & A   & {\color[HTML]{32CB00} Y}              & {\color[HTML]{32CB00} Y}                     \\ \hline
ArrivalLocation   & {\color[HTML]{32CB00} Y} & {\color[HTML]{32CB00} Y} & {\color[HTML]{32CB00} Y} & A                        & A                        & A   & {\color[HTML]{32CB00} Y}              & {\color[HTML]{32CB00} Y}                     \\ \hline
Trajectory        & {\color[HTML]{32CB00} Y} & {\color[HTML]{FD6864} N} & {\color[HTML]{FD6864} N} & {\color[HTML]{FD6864} N} & {\color[HTML]{FD6864} N} & {\color[HTML]{FD6864} N} & {\color[HTML]{32CB00} Y} & 
{\color[HTML]{FD6864} N} \\ \hline
\end{tabular}
\caption{Fields of datasets. Y stands for Yes, N stands for No, and A stands for Approximate. Though D4-D6 are subject to larger noises due to the limitation of crawling frequency (see Supplementary Note 2 for more details), the basic features regarding riding behaviours shown in Supplementary Figs. 1-2,4 are similar to other datasets provided by the platforms, which can partially prove that the quality of D4-D6 is acceptable.} 
\label{tab:dataFields}
\end{table}

\begin{table}[] \centering 
\begin{tabular}{|l|c|c|c|c|c|}
\hline
Shanghai                                    & \#trips         & $\langle d \rangle$ & gyration & revisitation & longest trip \\ \hline
(i)\ \  Random (non-spatial)                        & 0.2626          & 0.1214              & 0.6896   & 0.3732       & 0.1994       \\ \hline
(ii)\  Random (spatial, $r$=100 m)                      & \textbf{0.2500}          & \textbf{0.1157}              & \underline{0.3224}   & \textbf{0.2505}       & \textbf{0.1901}       \\ \hline
(iii)\  $times^{-1}$\ \  (top 10, $r$=100 m)           & 0.5165          & 0.1324              & 0.4549   & 0.3172       & 0.2692       \\ \hline
(iv)\  $times^{-1}\Delta t^{-1}$ (top 10, $r$=100 m)  & 0.3614          & 0.1747              & 0.3692   & 0.4097       & 0.2707       \\ \hline
(v)\ \  $times^{-1}\Delta t^{-1}$ (top 10, $r$=200 m) & 0.4207          & 0.2049              & 0.4169   & 0.5055       & 0.3329       \\ \hline
(vi)\  $times^{-1}\Delta t^{-1}$ ($r$=100 m)          & \underline{0.2557}          & \underline{0.1213}              & \textbf{0.3131}   & \underline{0.2674}       & \underline{0.1940}       \\ \hline
\end{tabular}
\caption{The KS distance between the generated distributions of bikes and the empirical one in Shanghai (D1). 
In Shanghai's case, the spatial proximity is quite important even a random selection within the searching range from the origin location of the trip can reproduce most distributions quite well, but it does not work very well in Beijing's case (see Table 1 in the main text) and Nanjing's case (see Supplementary Table 4). The choice model vi that incorporates the discovered scaling behaviour (the last row) outperforms other models on gyration and has the second smallest KS distance on all other indicators across cities. The smallest ones are highlighted in bold, and the second smallest are highlighted by underlines. 
} \label{tab:KS-shanghai}
\end{table}

\begin{table}[] \centering 
\begin{tabular}{|l|c|c|c|c|c|}
\hline
Nanjing                                    & \#trips & $\langle d \rangle$ & gyration & revisitation & longest trip \\ \hline
(i)\ \  Random (non-spatial)                        & 0.2301  & 0.0493              & 0.7865   & \textbf{0.0485}       & 0.0930        \\ \hline
(ii)\  Random (spatial, r=100 m)                            & 0.1426  & 0.0283              & 0.4737   & 0.3283       & 0.0614       \\ \hline
(iii) $times^{-1}$ (top10, $r$=100 m)               & 0.2623  & 0.0255              & 0.4795   & 0.3075       & 0.0949       \\ \hline
(iv)\  $times^{-1}\Delta t^{-1}$ (top10, $r$=100 m) & 0.1583  & 0.0418              & 0.4764   & 0.3595       & 0.0861       \\ \hline
(v)\ \  $times^{-1}\Delta t^{-1}$ (top10, $r$=250 m)  & \underline{0.0980}  & \underline{0.0214}              & \textbf{0.4209}   & \underline{0.1663}       & \underline{0.0380}       \\ \hline
(vi)\  $times^{-1}\Delta t^{-1}$ ($r$=250 m)        & \textbf{0.0693}  & \textbf{0.0167}              & \underline{0.4305}   & 0.3007       & \textbf{0.0236}       \\ \hline
\end{tabular}
\caption{The KS distance between the generated distributions of bikes and the empirical one in Nanjing (D3). 
In D3, as there is no user ID provided, thus we have to treat the user of each trip as a new individual and apply the discovered scaling behaviour in Fig. 4b in the main text. It is slightly strange that the most unrealistic non-spatial random selection works the best on the revisitation indicator (see the fourth column of the table). For all other indicators, the The choice model vi that incorporates the discovered scaling behaviour outperforms other models. Across all three cities (D1-D3), the choice model vi generally over-performs others (see Table 1 in the main text and Supplementary Table 3). The smallest ones are highlighted in bold, and the second smallest are highlighted by underlines. 
} \label{tab:KS-nanjing}
\end{table}

\begin{table} \centering
\begin{tabular}{|l|l|l|l|}
\hline
+bike-moving                                     & gyration (Shanghai) & gyration (Beijing) & gyration (Nanjing) \\ \hline
(ii)\ \ Random (spatial, $r$=100 m)                    & 0.2793           & 0.3606 & 0.3863             \\ \hline
(v)\ \,\,\,$times^{-1}\Delta t^{-1}$ (top 10)  & 0.2722 ($r$=100 m)           &   0.2774 ($r$=150 m)  & 0.2611 ($r$=250 m)         \\ \hline
(vi)\ \,$times^{-1}\Delta t^{-1}$          &  \textbf{0.2690} ($r$=100 m)   &  \textbf{0.2726} ($r$=150 m)  & \textbf{0.2574} ($r$=250 m)       \\ \hline

\end{tabular}
\caption{The KS distance between the generated distribution of bikes on gyration with an artificial bike-moving process and the empirical one in Shanghai, Beijing, and Nanjing (D1-D3). We can find that after calibrated by an even partial random moving process, the KS distance has greatly reduced as compared to the ones reported in Table \ref{tab.statistics} in the main text. In our multi-agent simulation system, after finishing each trip, a bike will be moved if the generated random number is smaller than 2/3 for Beijing and Nanjing (0.83 for Nanjing), then we extract a distance from the empirical bike-moving distance distribution (see Supplementary Fig. \ref{fig:moving_distance}) and then a random angle from 0$^\circ$ to 360$^\circ$. Taken together, we determine the next location of the bike. With such an artificial bike-moving process. Such results indicate that the bike-moving strategy during the period of the datasets might be no better than a partial random moving. The related distributions are shown in Supplementary Fig. \ref{fig:predictedBikeMobility}f,l,r.} \label{tab.KS-moving}
\end{table}

\subsection*{Supplementary Notes 1: Dataset Descriptions and Preprocessing}

To study the biking behaviours, mobility patterns of riders and sharing conveyances, and the relation between users and bikes in complex dockless bike-sharing systems, we exploit seven large-scale databases across cities with fine spatio-temporal resolution  (see brief summary in Supplementary Table \ref{tab.statistics}). These datasets cover seven highly diversified cities across countries: Shanghai (D1), Beijing (D2), Nanjing (D3-D4), Chengdu (D5), Xi'an (D6), Xiamen (D7), and Singapore (D8). The cities of D1-D7 are located in different regions of China. These cities are of diverse demographic and socio-economic status, urban geography, climate, and regional custom. 
Beijing and Shanghai are two mega cities with more than 20 million residents, while others are from around ten million down to a few million. These cities are far from each other and are located in different regions. 
The datasets span several years, from 2016 to the end of 2020. The time period also covers all four seasons, from spring to winter. For Nanjing, we obtain two datasets (D3 and D4) from two different platforms to make test if mobility patterns are consistent across platforms or not, as different platforms have some difference on operation and promotion strategy. D5 and D6 are of the same time period, and obtained from two different cities. D7 is a aggregated datasets from the three biggest platforms in Xiamen, and only provides data during the morning rush hours. 

D1 is obtained from the \textit{Mobike} platform. It captures 1.02 million biking trips of 17 thousand users, which involves 0.3 million dock-less sharing bikes over one month (from 08/01/2016 to 09/01/2016) in Shanghai. Note that the dockless bike-sharing service was officially launched on 04/22/2016 in China for the first time\footnote{The trial operation of Mobike in Shanghai was started from 12/06/2015. In order to make the bike more durable and easier to be maintained, the design of dockless sharing bikes of Mobike adopted a shaft drive instead of a common chain drive. Ofo, founded in the campus of Peking University in April 2014, was actually the first dockless bike-sharing platform that appeared in China, whose service was launched on 09/07/2015, which is earlier than the Mobike.  However, Ofo was restricted to the campus of universities and only started trial operations in the urban area of Beijing and Shanghai in late October, 2016.}, the Shanghai dataset (D1) starts roughly 3 months after the launching of dockless sharing bikes, so back then dockless bike is still a new thing, people might be quite curious about it and enthusiastic to try it out at that time; in addition, during the early promotion stage, the \textit{Mobike} company also had great special offers to users. In comparison, other datasets are one or more years after the launching of dockless sharing bikes (so users might be quite used to it). For example, Mobike entered Beijing on 08/16/2016, thus D2 is roughly one year after its service launching in Beijing. Mobike Later entered a few other first-tier and second-tier big cities in China in 2016, including Guangzhou (10/27/2016), Shenzhen (11/19/2016), and Chengdu (in late November 2016).   

In addition, the information regarding the trip in D1 is the most detailed among all datasets (see Supplementary Table \ref{tab:dataFields}).  
Each record has an order ID, a user ID, a bike ID, departure and arrival locations, and their corresponding timestamps, as well as the pass-by points of the trajectory (see Supplementary Table \ref{tab:dataFields}). This enables us to make accurate calculations on the riding distance, riding duration, and average riding speed of each trip.  
We assume that the real riding distance is the shortest path connecting all pass-by points. Then the average riding speed of each trip in D1 can be calculated. 
For other datasets (D2-D6), there is no trajectory, thus accurate riding distance and riding speed can not be calculated; in D7, we find that trajectories are labeled with another set of ids that can not be matched with the ids of riding trips.

D2 is also a dataset obtained from the \textit{Mobike} platform. It has more than 3.2 million records, detailing the riding behaviours of 0.35 million users and 0.485 million dockless sharing bikes over two weeks (from 05/10/2017 to 05/24/2017) in Beijing. In D2, the arrival time of the trip is not provided. Though we can obtain the routing distance and expected travel time via Amap API, the travel time is still an estimation and we did not report the distribution of riding duration of the Beijing dataset in Supplementary Fig. \ref{fig:waitingInterval_duration}. In addition, the Amap API can also return the routing trajectory between any two locations, thus the trajectories in Fig. 1f,h in the main text are obtained in this way.

D3 is obtained from another big dockless bike-sharing platform \textit{DiDi Bike} in China. It has 1.35 million records over one month (from 10/01/2019 to 10/31/2019) recording the trips of 0.06 million dockless sharing bikes. Note that hashed user ID is not provided in the Nanjing dataset (D3) and other datasets (D4-D8), thus user-related mobility patterns in Fig.~2 in the main text cannot be analysed. 

D4-D6 are crawled online from the \textit{Mobike} platform. The crawling algorithm works in the following way: at each time stamp, the API will return information about those available bikes, while the bikes that are in use cannot be detected. Due to the limitation on computing resources, the crawling frequency is only guaranteed to be less than 10 minutes. The raw data we obtained is ``a time, the location of bikes, and the bike ID''. When a bike is ``disappeared'' in the next time slot, it would be highly probably picked up by a user (yet the user ID is not detectable, as we cannot obtain any information regarding ongoing trips), and the last location of the bike will be regarded as the departure location, and the last time is the approximate departure time. When the bike ``appears'' again in a future time slot, we identify the new location as its arrival location of the last trip and the time as the approximate arrival time. 
So in this way, we can roughly reconstruct a dataset similar to D3. Yet the departure (and arrival) time and location is only a rough approximation, especially on the timestamps (see Supplementary Table \ref{tab:dataFields}). If two trips happen within 10 minutes, we might miss such trips due to the limited crawling frequency (especially in the case of D5, see Supplementary Fig. \ref{fig:waitingInterval_duration}c). 
To avoid further stronger noises, after obtaining the raw data, we identify a trip as a displacement longer than 100 meters and less than 120 minutes between two consecutive records. While bike-moving due to rebalancing is harder to detect in these datasets. We find that the distribution of riding duration of D4 and D6 is quite close to the ones in other cities, only D5 is relatively different at the head part due to the crawling frequency limitation (see Supplementary Fig. \ref{fig:waitingInterval_duration}c). As for the waiting interval between two consecutive trips of the same bike, the distributions of D4-D6 are quite close to other cities (see Supplementary Fig. \ref{fig:waitingInterval_duration}d). Results presented in Supplementary Fig. \ref{fig:waitingInterval_duration}b-d indicate that the quality of D4-D6 should be acceptable.  

The main consideration of crawling data for Nanjing from the \textit{Mobike} (D4) is to make comparisons between different dockless bike-sharing platforms and different time periods. The results in the main text and the supplementary materials show that they are qualitatively the same, and quantitatively similar (see Figs. 2-4 in the main text and Supplementary Figs. 1,2,4). D5 and D6 are of the same time period to make better comparisons between cities.  

D7 is aggregated from the three biggest platforms in Xiamen -- Meituan Bike (formerly Mobike)\footnote{Since January 23, 2019, Mobike changed its name as Meituan Bike as part of an integration with its parent company Meituan Dianping.}, DiDi Bike, and Hellobike. Note that D7 only has data during the morning rush hours from 6 a.m. to 10 a.m. (four hours of each day), not whole days as in D1-D6. The raw data in D7 has two parts. One is the order dataset which contains ``an order id, a bike id, a timestamp, location, and the status of the bike'', the status is either ``open'' or ``locked'', thus by connecting two records of the same bike between open status to locked status, we can identify the biking trips and format the data as of D1-D3. The other one is the trajectory dataset, but the ids are different from the ones in the order dataset, thus these two datasets cannot be matched together easily, and we did not use this trajectory dataset in this work.  

D8 is from a dockless bike-sharing operator in Singapore, which has 0.298 million records over one week (from 09/11/2017 to 09/17/2017) recording the trips of 0.033 million dockless sharing bikes. The dataset is the one shared by Ref.\cite{kondor2021estimating}. More details regarding data collection and preprocessing procedures of D8 were presented in the supplementary materials of Ref.\cite{kondor2021estimating} and Refs.\cite{shen2018understanding,xu2019unravelBIKE}.

\subsubsection*{The travel distance of each trip}
In the Shanghai Dataset (D1), the pass-by points of the biking trajectory of each trip are unsorted (without timestamp neither), thus we assume that the shortest path that connects all of the pass-by points is the real trajectory of the trip. Finding the shortest path is implemented by the ant colony optimization algorithm \cite{dorigo2006ant}. The travel distance distribution is reported in Fig. 1a in the main text. 
We also compare it with the distribution of Euclidean distance between the origin and destination of each trip and find it significantly deviates from the Euclidean one at the head part as well as the tail part, i.e., for the short trips that are less than 1 km and long trips that are larger than 10 km (see Supplementary Fig.~\ref{fig:waitingInterval_duration}a).  

In the Beijing dataset (D2), there are no pass-by points of the trajectory, so we obtain the routing distance and riding time between any two locations from the Amap Open Platform (\url{https://lbs.amap.com/}), whose API is open to everyone. The route planning API takes the real-time complex conditions of road networks (including the over-bridges, one-way roads, and road closures) into considerations. In the Amap, we find that vehicle traffic almost has no impact on the travel time by bike, the speed of biking is roughly identical under different traffic conditions. 
We choose a speed-priority route among all of the optional routes, and get its distance and estimated travel time (and the turning points along with the route for the cases in Fig. 1f,h in the main text). We discover that the routing distance is quite close to the Euclidean distance in both Beijing and Shanghai (see Supplementary Fig. ~\ref{fig:waitingInterval_duration}a). As the actual arrival timestamp is not provided in D2 (see Supplementary Table \ref{tab:dataFields}), though we can estimate the arrival time via Amap API based on the origin-destination (OD) locations of the trip, it is still not the real case and thus we did not report its distribution in Supplementary Fig.~\ref{fig:waitingInterval_duration}c. 

As in other datasets (D3-D8), in which there is also no trajectory provided (or cannot be matched in D7), we only report the Euclidean straight-line distance distribution in Supplementary Fig.~\ref{fig:waitingInterval_duration}b. The distribution of Euclidean straight-line distance across cities can be reasonably well fitted by a truncated power-law $P(d)=(d+d_0)^{-\alpha} e^{-d/\kappa}$, where $d$ is the displacement distance, and $\kappa$ is the cut-off distance after which the probability would drop dramatically (see Supplementary Fig. 1b). This observation is qualitatively consistent with previous findings \cite{brockmann2006scaling,gonzalez2008understanding,alessandretti2017multi,barbosa2018human}, yet the scaling exponents ($\alpha=$3.47) is much larger than the ones of all travel modes, which is usually between 1 and 2 \cite{alessandretti2017multi}. This indicates that biking trips have a much faster decay along with the increase of travel distance than other means of transportation. 



\subsection*{Supplementary Note 2: Multi-agent simulation platform of the dockless bike-sharing system}
In order to test the impacts of different choice models on the mobility patterns of sharing bikes, we build a multi-agent simulation platform, in which the users and bikes are initialized with the location when he/she/it first appeared. The real spatio-temporal travel demands are streamed to the simulation platform, we simulate the same origin-destination of the trips in reality, but bikes will be chosen according to several human choice models, and thus the real-time distribution of bikes in the simulation system will not be the same with the real situation. By tracking every move of each bike in the simulation system, we can generate related distributions of bikes on mobility patterns when the simulation ends. 

We first start with the most unrealistic but most random scenario (referred as ``choice model (i) Random (non-spatial)'' in Table 1 in the main text and Supplementary Table \ref{tab:KS-shanghai}): we allow the user in the simulation platform to randomly choose any bike in the city regardless of the spatial position of it, and then the trip will be associated with the chosen bike. After fulfilling all travel demands of users, every trip will be assigned to a bike, and then we make mobility statistics on bikes to generate related distributions and compare them with the empirical one to get the corresponding KS distance. 

Then we become more realistic to consider the spatial constraints that a user can only pick up a bike within a certain searching range (e.g., a radius $r$ equals to 100 meters) from him/her origin of the trip (referred as ``choice model (ii) Random (spatial, $r$=100 m)''): we find that on some indicators, the KS distance becomes smaller in the case of Shanghai (see Supplementary Table \ref{tab:KS-shanghai}), but, in the case of Beijing, some even become larger in Table 1 in the main text, which indicates that spatial constraint is not the only factor that influences the choice behaviour. 

Furthermore, we assume that the condition of a bike would also have an impact, so we calculate the number of trips (marked as $times$) that a bike has been taken until now, as well as the standby time of the bike since its last arrival (marked as $\Delta t$). A larger number of trips would usually mean more serious wear of the bike itself or even a higher probability of ridden by a reckless user who does not care for the bike. Longer standby time $\Delta t$ would usually correspond to a higher probability of collecting dust or even bird droppings, which makes it much less attractive to riders. So a smaller $times$ or $\Delta t$ would highly probability corresponds to a newer and better bike. 
Then we first take the number of trips into consideration (referred as ``choice model (iii) $times^{-1}$ (top 10, $r$=100 m)''): within a certain searching range ($r$=100 m), the user will randomly choose one bike out of the top 10 ranked bikes based the $times^{-1}$ indicator. Again, the KS distance on some indicators becomes smaller, but some become larger. 
We then further take the standby time $\Delta^{-1}$ into consideration and come up with the composite indicator $times^{-1} \Delta^{-1}$ (choice model iv), and discover that the KS distance on some indicators become better and some slightly worse (see Table 1 in the main text and Supplementary Tables 3-4).

We assume that the searching range also plays an important role, and we find that when further enlarge the searching range $r$ in Beijing and Nanjing (choice model v), the KS become much smaller than most of the previous choice models (see the fourth and fifth row in Table 1 in the main text and Supplementary Table 4). But this is not the case in Shanghai, after enlarging the searching range $r$, the KS distance all become larger (see the fourth and fifth row in Supplementary Table \ref{tab:KS-shanghai}). This inspires us that the characteristic spatial scale of two cities might be different, which motivates us to make more comprehensive studies on the effect of the searching scale $r$ as reported in Fig. 4a in the main text and Supplementary Fig. \ref{fig:numberDistribution}. 
In addition, apart from random selection or random selection from the top ten bikes, we wonder what is the real distribution on the rank of the selected bike in reality, and we surprisingly discover an emergent behaviour, which is reported in Fig. 4b in the main text and Supplementary Fig. \ref{fig:rescaledRankDistribution}. 
We then propose the ``choice model vi'' that incorporates the discovered scaling behaviour ($P(rank)\propto rank^{-\alpha} e^{-rank/\kappa}$), and find that the KS distance becomes much smaller on most indicators across cities (see Table 1 in the main text and Supplementary Tables 3-4). This indicates that the spatial scale together with the emergent scaling behaviour can well explain the distribution of bikes on mobility despite great diversities across cities. The cities we studied (Beijing, Shanghai, Nanjing, Chengdu, Xi'an, and Xiamen) have quite different characteristics, including 
different size of the city, user-bike ratios (see Supplementary Table 1), different characteristics of public bicycling infrastructure, road networks and urban terrain, different climate, and weather, promotion strategy, different level of curiosity on sharing bikes at the time of the datasets, or even different attitudes towards biking induced by culture and custom.

\end{document}